\documentclass[11pt]{article}
\usepackage[preprint]{tmlr}

\usepackage{amsmath,amsfonts,bm}

\def\eqref#1{equation~\ref{#1}}

\def\1{\bm{1}}

\DeclareMathAlphabet{\mathsfit}{\encodingdefault}{\sfdefault}{m}{sl}
\SetMathAlphabet{\mathsfit}{bold}{\encodingdefault}{\sfdefault}{bx}{n}

\newcommand{\E}{\mathbb{E}}

\usepackage[utf8]{inputenc}
\usepackage[T1]{fontenc}
\usepackage{lmodern}
\usepackage[margin=1in]{geometry}
\usepackage{amsmath,amssymb,amsthm}
\usepackage{graphicx}
\usepackage{booktabs}
\usepackage{array}
\usepackage{multirow}
\usepackage{xcolor}
\usepackage{microtype}
\usepackage{enumitem}
\usepackage{caption}
\usepackage{subcaption}
\usepackage{pgfplots}
\pgfplotsset{compat=1.17}
\usepgfplotslibrary{groupplots}
\usepackage[hidelinks]{hyperref}
\usepackage{cleveref}
\usepackage{authblk}

\let\cite\citep

\theoremstyle{definition}
\newtheorem{definition}{Definition}
\newtheorem{hypothesis}{Hypothesis}
\newtheorem{assumption}{Assumption}
\theoremstyle{plain}
\newtheorem{proposition}{Proposition}

\theoremstyle{remark}
\newtheorem{remark}{Remark}

\newcommand{\Prob}{\mathbb{P}}

\definecolor{c0}{HTML}{1f77b4}
\definecolor{c1}{HTML}{d62728}
\definecolor{c2}{HTML}{2ca02c}
\definecolor{c3}{HTML}{9467bd}
\definecolor{c4}{HTML}{ff7f0e}
\definecolor{c5}{HTML}{8c564b}
\pgfplotsset{
  every axis/.append style={
    label style={font=\small},
    tick label style={font=\footnotesize},
    legend style={font=\footnotesize},
    title style={font=\small\bfseries},
  },
  diaglines/.style={thick, mark=*, mark size=1.3pt},
}

\newcommand{\Phimax}{\Phi_{\mathrm{gcc}}}
\newcommand{\fone}{F_1}

\title{\bfseries Diagnosis as Percolation on an Evidence Graph:\\
Root-Cause Recoverability in Network Fault Diagnosis}

\author{\name Jerome Henry \email jerhenry@cisco.com

      Cisco Systems
      \AND
      \name Swadhin Pradhan \email swapradh@cisco.com
      
      Cisco Systems
      \AND
      \name Amine Choukir \email amchouki@cisco.com

      Cisco Systems
}

\begin{document}
\maketitle

\begin{abstract}
Automated diagnosis under partial observation recurs across computing and
engineering: an agent must infer an unobserved root cause from an incomplete set
of symptoms, and sometimes gauge how far to trust that inference. We argue that
\emph{random-graph theory} supplies both a model and a set of operational
primitives to solve this problem. We cast diagnosis as \emph{percolation on a partially observed
evidence graph}, whose vertices are observations, entities, hypotheses and
candidate causes and whose edges are derived deterministically from data; a
diagnosis is then a connected explanatory subgraph, and recoverability
corresponds to whether the evidence has \emph{percolated} into one giant cluster
rather than scattering into fragments. Our primary contribution is to
\emph{network fault diagnosis}, in the lineage of the \emph{PLUME}~\cite{pradhan2026plume} and \emph{PROBE}~\cite{henry2026probe} systems: we
work the abstraction out in full on an ensemble of large language models (LLMs)
that diagnoses faults from textualized 802.11 packet captures, where an
ensemble-plus-reconciliation pipeline lifts weighted evidence $\fone$ from
$\approx\!0.85$ to $\approx\!0.96$ yet provides no calibrated notion of trust.
Probing whether connectivity supplies that missing signal yields a cautionary
result: post-hoc connectivity correlates with correctness (Spearman $\rho$ up to
$0.67$), but under capture-clustered controls, degree-preserving negative-control
nulls and multiple-comparison correction, no structural signal retains value
beyond capture size. Connectivity is thus a portable, theoretically grounded
substrate for fault \emph{recoverability} and root-cause \emph{localization}, but
not a reliable proxy for answer-level trust. A controlled benchmark recovers the
predicted critical density $\phi_c\approx 1/c$, where a connectivity-aware
personalized-PageRank localizer tracks a Bayesian oracle while degree heuristics
collapse under decoy hubs; and studies in five further domains, with a
substrate-independent threshold analysis, indicate that the percolation
\emph{principle} transfers beyond networking.
\end{abstract}

\section{Introduction}
\label{sec:intro}

\subsection{Automated diagnosis under partial observation}
\label{ssec:general}
Diagnosis (inferring an unobserved cause from observed effects) is a
foundational task across computing, engineering and medical sciences. For example, a network operator infers a misconfiguration from a packet trace; a site-reliability engineer localizes the
failing microservice behind a cascade of alarms; a grid operator anticipates
which transmission lines will cascade; a clinician reasons from a set of
symptoms to a disease. These problems share three structural features that make
them difficult and that motivate a unified treatment.

\emph{First, they are abductive and many-to-one:} numerous latent causes can
produce overlapping observable effects, so an effect set rarely identifies a
unique cause. \emph{Second, observation is partial and non-uniform:} only a
fraction of the relevant evidence is ever recorded (a monitoring buffer
captures a short window, telemetry is sampled, a patient reports a subset of
symptoms), so the diagnostic substrate is a \emph{partially observed} object
rather than a complete one. \emph{Third, and increasingly, the diagnostic agent
must be trustworthy:} as machine-learning systems and large language models
(LLMs) are deployed to produce diagnoses automatically, a calibrated estimate of
\emph{when to trust} a diagnosis is as important as the diagnosis itself, since
an over-confident wrong answer is more harmful than an abstention.

We contend that these features are precisely those studied by \emph{random-graph
theory} and \emph{percolation}~\cite{erdos1960evolution,newman2003structure}.
If the evidence for a case is organized as a graph whose vertices are
observations, entities, symptoms, hypotheses, and candidate causes, then a
diagnosis is a \emph{connected explanatory subgraph} linking observed effects to a
cause, and the central percolation question ``has a giant connected component
emerged?'' becomes a question about recoverability: if the observed evidence
\emph{percolates} into one connected cluster reaching a candidate cause, the cause
is \emph{recoverable}; if it stays fragmented, no method can confidently identify
it. This paper develops that correspondence into a model, a family of computable
signals, and a body of cross-domain evidence---and, equally importantly, maps its
boundary: connectivity is \emph{not} a reliable proxy for whether a given answer is
\emph{trustworthy} once observed-evidence volume is accounted for
(\Cref{sec:signals}).

\subsection{A demanding instance: LLM diagnosis of packet captures}
\label{ssec:pcap}
We ground the abstraction on a concrete and unforgiving instance: diagnosing
faults directly from IEEE~802.11 packet captures (PCAPs). A monitoring access
point retains a rolling buffer of recent frames; when an anomaly is detected,
the buffer is exported for analysis. Two complementary systems define the
state of the art we build upon. \emph{PLUME} is a network-native foundation
model that learns representations of 802.11 traces through protocol-aware
tokenization~\cite{pradhan2026plume}; \emph{PROBE} is a multi-stage,
evidence-grounded ensemble pipeline that renders a textualized PCAP into a
frame-cited diagnosis accompanied by a deterministic reliability
score~\cite{henry2026probe}. Both expose each capture to an LLM and ask it to
identify the fault; both are evaluated against a \emph{golden set} in which
human experts record a chain-of-thought~\cite{wei2022chain} and a structured
verdict (the issue, the responsible entity, and the citing frames), enabling
automatic scoring by a weighted evidence-overlap $\fone$.

This instance is demanding for two reasons. A single LLM pass is unreliable:
instruction-tuned models are post-trained toward caution and tend to commit only
when an explicit error message is present, missing subtle evidence distributed
across many frames. An ensemble of stochastic drafts followed by a
\emph{reconciliation} step recovers much of the lost signal, an instance of the
variance reduction underlying self-consistency
decoding~\cite{wang2023selfconsistency} and the classical bias--variance
decomposition~\cite{geman1992bias}. \Cref{tab:ablation} reproduces the headline
ablation from \emph{PROBE}~\cite{henry2026probe}: the
ensemble-plus-reconciliation configuration~(C) attains weighted evidence
$\fone = 0.964$, versus $0.842$--$0.871$ for leaner variants. Reconciliation is
thus the dominant accuracy lever but, as we discuss next, it leaves the
question of \emph{trust} unanswered.


\subsection{The reliability gap left open by reconciliation}
\label{ssec:gap}
Reconciliation determines \emph{what} the consensus diagnosis is, but not
\emph{how much it should be trusted}, and the latter is essential for
automated deployment. Two natural sources of confidence prove inadequate. The
model's \emph{self-reported} confidence is uninformative: $74$ of $104$ golden
cases report a confidence of exactly $0.95$, and stratifying accuracy by
self-reported confidence yields a non-monotone curve (\Cref{fig:calib}), the
familiar miscalibration of modern neural
predictors~\cite{guo2017calibration}. A hand-weighted deterministic composite
of run-to-run stability and evidence-validity terms is better behaved (accuracy
increases monotonically with the score) but its weights were chosen manually
and it lacks a principled, structural foundation. We therefore seek a
reliability signal that is (a) \emph{intrinsic} to the evidence a diagnosis
cites, (b) \emph{deterministic} and inexpensive to compute, and (c)
\emph{theoretically grounded} in a model of why connected evidence indicates a
reliable inference.

\subsection{Contributions}
\label{ssec:contrib}
\begin{enumerate}[leftmargin=1.4em,itemsep=2pt]
\item \textbf{A percolation model of diagnosis} (\Cref{sec:model}): an evidence
graph with a generative partial-observation process, four connectivity
\emph{order parameters} grounded in generating-function percolation theory and
personalized-PageRank relevance, and a bias result (\Cref{prop:bias}) explaining
why a \emph{closed-world} noisy-OR posterior is fragile under under-observation---a
misspecification of the observation process, not a failure of Bayes.
\item \textbf{A falsification of the bridging hypothesis} (\Cref{sec:bridging}):
on $200$ real reconciliation pairs, repair does \emph{not} significantly raise the
giant-component fraction; its measurable effect is \emph{orphan pruning}
($0.149\!\to\!0.069$, $p{=}0.008$), while post-reconciliation connectivity
correlates with correctness ($\rho$ up to $0.67$)---an association we trace
largely to capture size.
\item \textbf{A rigorous negative result on structural reliability signals}
(\Cref{sec:signals}): across six candidate signals, with capture-clustered
inference, degree-preserving nulls, and multiple-comparison correction,
\emph{none} retains value once capture size and trivial citation statistics are
controlled---the apparent ``structure-as-trust'' signal is a capture-size artifact.
\item \textbf{A controlled percolation benchmark} (\Cref{sec:bench}) whose
empirical critical density approaches $\phi_c\approx1/c$ and isolates an
\emph{algorithmic detectability gap}: a connectivity-aware localizer tracks a
Bayesian oracle while degree-only heuristics collapse.
\item \textbf{Cross-domain generality} (\Cref{sec:cross}): a reverse-PageRank
localizer attains $89\%$ top-1 on a \emph{real} microservice chaos corpus (Train
Ticket); power-grid cascades and three controlled simulations (software fault
localization, multi-hop RAG, security attack-path) reproduce the
recoverability/localization result; and bipartite, biased-observation, and
phase-boundary analyses show the threshold itself is substrate-independent. We
claim transfer of the \emph{principle}, not of a tuned system.
\end{enumerate}
All experiments are deterministic given a random seed and use only public data.

\paragraph{Scope and positioning.}
This is first a paper about \emph{network} fault diagnosis, in the lineage of
\emph{PLUME}~\cite{pradhan2026plume} and \emph{PROBE}~\cite{henry2026probe}: our real ground
truth and primary claims concern 802.11 packet-capture diagnosis. The
cross-domain studies (\Cref{sec:cross}) show the percolation \emph{principle} is
substrate-independent---a property inherited from the mathematics, not from any one
dataset---and we are explicit about which findings rest on \emph{real} data and
which are \emph{controlled simulations}.

\section{Background and Related Work}
\label{sec:bg}

We summarize the theory we draw on, stating the results in the form we will use
them, and then position the present work against prior literature on LLM
reliability and cross-domain diagnosis.

\paragraph{Random graphs and percolation.}
The Erd\H{o}s--R\'enyi ensemble $G(n,p)$ exhibits a phase transition: a giant
connected component (GCC) of size $\Theta(n)$ emerges once the mean degree
exceeds unity~\cite{erdos1960evolution}. For a general degree distribution the
standard generating-function derivation (\Cref{app:gf}) gives the
bond-percolation threshold
\begin{equation}
\phi_c=\frac{1}{G_1'(1)}=\frac{\langle k\rangle}{\langle k^2\rangle-\langle k\rangle},
\label{eq:phic}
\end{equation}
where $G_1$ is the excess-degree generating function and $\langle k\rangle$,
$\langle k^2\rangle$ are the first two degree moments; for a Poisson degree
distribution of mean $c$ this is the clean prediction $\phi_c=1/c$ tested in
\Cref{sec:bench}. Two consequences recur throughout: near $\phi_c$ the
\emph{susceptibility} $\chi$ (mean size of the non-giant component containing a
random vertex) diverges, giving a finite-sample locator of the transition; and
\Cref{eq:phic} explains the robustness asymmetry---deleting vertices in
descending centrality order maximally reduces $\langle k^2\rangle$ and destroys
the GCC far faster than random deletion~\cite{albert2000error,
cohen2000resilience}, a property we exploit as a diagnostic in \Cref{sec:cross}.

\paragraph{Random walks, centrality, and relevance.}
PageRank ranks vertices by the stationary distribution of a restarting random
walk~\cite{page1999pagerank}; its personalized variant~\cite{haveliwala2002topic}
restarts from a query set $F$ and so measures \emph{relevance to $F$}, which we use
as a domain-agnostic prior over candidate causes. Edge betweenness
centrality~\cite{brandes2001faster} identifies structural bottlenecks and ranks
lines likely to carry power-grid cascades. Formal definitions are in
\Cref{ssec:relevance}.

\paragraph{Probabilistic diagnosis and the noisy-OR model.}
Abductive diagnosis has a long history in probabilistic
reasoning~\cite{pearl1988probabilistic}. The noisy-OR model and
QMR-DT~\cite{shwe1991probabilistic} assume each cause independently activates each
finding, yielding a tractable likelihood whose exact posterior we use as a
Bayesian \emph{oracle}; \Cref{prop:bias} shows its treatment of \emph{absent}
findings is what makes it fragile under partial observation, motivating
positive-evidence structural priors.

\paragraph{LLM reliability, calibration, and selective prediction.}
Self-consistency decoding~\cite{wang2023selfconsistency},
retrieval-augmentation~\cite{lewis2020retrieval}, and chain-of-thought
prompting~\cite{wei2022chain} reduce error and elicit citable reasoning; the
ensemble-plus-reconciliation pipeline we build on~\cite{henry2026probe} is in this
lineage. Because neural predictors are typically
miscalibrated~\cite{guo2017calibration}, a separate intrinsic trust signal is
needed; we adopt selective
classification~\cite{elyaniv2010foundations,geifman2017selective} and its
risk--coverage curve as our yardstick (\Cref{ssec:metrics}).

\paragraph{Cross-domain diagnosis benchmarks.}
We evaluate portability on problems with distinct graph topologies: microservice
RCA on directed service-call graphs (Train Ticket~\cite{zhou2021trainticket}),
power-grid cascading failure on undirected bus--branch networks
(PowerGraph~\cite{varbella2024powergraph}), and bipartite QMR-DT
diagnosis~\cite{shwe1991probabilistic}, synthetic and
real~\cite{itachi2019dataset,wang2008automated}. Treating these with a single
structural prior is, to our knowledge, novel.

\section{Diagnosis as Percolation on an Evidence Graph}
\label{sec:model}

\subsection{The evidence graph and its observation process}
\label{ssec:evgraph}
\begin{definition}[Evidence graph]
\label{def:evgraph}
For a single case, let the vertex set $V = O \cup E \cup S \cup H \cup R$
comprise \emph{observations} $O$ (e.g.\ packet frames, log lines, telemetry
points), \emph{entities} $E$ (endpoints, services, components), \emph{symptoms}
$S$, \emph{hypotheses} $H$, and candidate \emph{root causes} $R$. The
\emph{latent} evidence graph $G^\star=(V,A^\star)$ has an edge set $A^\star$
generated deterministically by domain rules of three kinds: \emph{temporal}
adjacency between observations, \emph{co-membership} in a shared
conversation/entity, and \emph{citation} links connecting a diagnosis to the
observations and entities it references. A diagnosis that cites a vertex set
$C\subseteq V$ induces the \emph{cited subgraph} $G[C]$, the subgraph of
$G^\star$ on $C$.
\end{definition}

Two design choices make \Cref{def:evgraph} portable. First, the edge rules are
\emph{deterministic} functions of the data, so the graph, and every quantity
derived from it, is reproducible and independent of the diagnostic model.
Second, the rules are \emph{local}: each requires only adjacency or shared
membership, which every domain we study supplies. In the PCAP instance the
pipeline exposes only the rendered text and the set of cited frame numbers, so
we use a \emph{time-and-conversation} rule: cited frames $i,j$ are adjacent iff
$|i-j|\le 1$ in capture order (a unit-width temporal bridge) or they belong to
the same endpoint conversation. \Cref{sec:cross} instantiates the same
definition on directed service graphs, undirected bus--branch networks, and
bipartite cause--finding graphs.

We model partial observability explicitly, since it is the crux of the problem
(\Cref{ssec:general}).
\begin{assumption}[Partial observation]
\label{ass:obs}
The diagnostic agent does not see $G^\star$ but a random subgraph $G_\phi$
obtained by retaining each observation independently with probability $\phi\in
(0,1]$ and keeping an edge iff both endpoints are retained. Equivalently,
$G_\phi$ is a site-percolated copy of $G^\star$ with occupation density $\phi$.
\end{assumption}
\Cref{ass:obs} is a missing-completely-at-random idealization of buffered
capture, telemetry sampling, and incomplete symptom reporting. It is the bridge
between the engineering problem and percolation theory: the visible evidence is
literally a percolated graph, so the percolation threshold $\phi_c$ of
\Cref{eq:phic} predicts the observation density below which the evidence
fragments and diagnosis becomes information-theoretically hard.

\subsection{Connectivity order parameters}
\label{ssec:orderparams}
We summarize the connectivity of a cited subgraph $G[C]$ by four scalar
\emph{order parameters}:
\begin{align}
\Phimax(C) &= \frac{|\text{largest connected component of } G[C]|}{|C|}
  &&\text{(giant-component fraction)},\label{eq:phigcc}\\
\kappa(C) &= \#\{\text{connected components of } G[C]\}
  &&\text{(fragmentation)},\\
\iota(C) &= \frac{\#\{\text{isolated vertices of } G[C]\}}{|C|}
  &&\text{(orphan fraction)},\\
\delta(C) &= \frac{2\,|A(G[C])|}{|C|\,(|C|-1)}
  &&\text{(edge / span density)}.
\end{align}
$\Phimax$ is the finite-sample estimator of the percolation order parameter
$S=\lim_{n\to\infty}|\text{GCC}|/n$, the largest root in $[0,1]$ of
$S = 1-G_0(1-S\,\phi)$ under \Cref{ass:obs}~\cite{newman2003structure}. The
fragmentation $\kappa$ and orphan fraction $\iota$ measure sub-criticality from
complementary angles: $\kappa$ counts the components a correct explanation
should fuse, while $\iota$ isolates the ``dangling'' evidence a diagnosis cited
but failed to connect to anything else. The density $\delta$ is the realized
fraction of possible internal edges and behaves, empirically, as the most
informative single summary (\Cref{sec:bridging}).

\subsection{The percolation hypothesis}
\label{ssec:perchyp}
\begin{hypothesis}[Diagnosis-as-percolation]
\label{hyp:perc}
A trustworthy diagnosis corresponds to a cited subgraph that has
\emph{percolated}: a single explanatory component spans the relevant evidence
($\Phimax\!\to\!1$, $\kappa\!\to\!1$, $\iota\!\to\!0$). Under-observation
($\phi<\phi_c$) or an incorrect diagnosis leaves the subgraph
\emph{sub-critical}: fragmented, with many orphans.
\end{hypothesis}
\Cref{hyp:perc} has a precise prediction under
\Cref{ass:obs,eq:phic}: as $\phi$ crosses $\phi_c$ from below, the probability
that the observed evidence supports a connected explanation rises sharply from
near zero to order one, with a transition that sharpens as the system grows.
\Cref{sec:bench} confirms this prediction in a controlled benchmark, and
\Cref{sec:bridging} tests its corollary that more-connected cited subgraphs are
more often correct.

\subsection{Relevance as a damped random walk}
\label{ssec:relevance}
The central localization primitive is \emph{personalized
PageRank} (PPR)~\cite{page1999pagerank,haveliwala2002topic}, seeded at the
observed evidence set $F\subseteq V$. With damping $d\in(0,1)$, restart
distribution $\mathbf{s}$ uniform on $F$, and $P$ the row-stochastic adjacency
of $G_\phi$, the relevance vector $\pi$ is the unique solution of
\begin{equation}
\pi = (1-d)\,\mathbf{s} + d\,P^\top \pi .
\label{eq:ppr}
\end{equation}
Because $d\,P^\top$ has spectral radius $d<1$, \Cref{eq:ppr} admits the
convergent Neumann-series solution
\begin{equation}
\pi = (1-d)\sum_{t\ge 0} d^{\,t}\,(P^\top)^t\,\mathbf{s},
\label{eq:ppr-series}
\end{equation}
which exposes the prior's meaning: $\pi_v$ is a length- and multiplicity-weighted
count of walks from the observed evidence to $v$, discounted by $d^t$ with path
length $t$. A candidate cause is therefore scored by how \emph{reachably
relevant} it is to the present evidence, a quantity that ignores absent
evidence entirely. This property is what underlies the robustness result below.
Candidate causes are ranked by $\pi$; when causes are known to lie
\emph{upstream} of effects (directed graphs), we transpose the walk, which is
decisive for microservice RCA (\Cref{sec:cross}).

\subsection{The noisy-OR posterior and its closed-world fragility}
\label{ssec:noisyor}
In bipartite cause$\to$finding settings (the QMR
tradition~\cite{shwe1991probabilistic,pearl1988probabilistic}), the exact
single-cause noisy-OR posterior (\Cref{eq:noisyor}, \Cref{app:noisyor}) is a
Bayesian \emph{oracle} at full observation. Its \emph{closed-world} use under
partial observation---scoring every unreported finding as \emph{absent}---is the
source of its fragility: under \Cref{ass:obs} a finding generated by the true
cause but not observed is treated as negative evidence, penalizing the true cause
for a symptom it actually produced. \Cref{prop:bias} makes this precise; it is a
misspecification of the observation process, not a failure of Bayes.

\begin{proposition}[Missingness bias of the closed-world noisy-OR]
\label{prop:bias}
Fix a true cause $r^\star$ generating finding set $F^\star$, let each finding in
$F^\star$ be observed independently with probability $\phi$ (\Cref{ass:obs}), and
write $p_f := \Prob[f{=}1\mid r^\star]$ for the per-finding activation
probability. If unobserved findings are scored as \emph{absent} (the closed-world
reading of \Cref{eq:noisyor}), the expected change in the log-likelihood of
$r^\star$ relative to full observation is
\begin{equation}
\E\big[\log\Prob_{\mathrm{cw}}[F_\phi\mid r^\star]-\log\Prob[F^\star\mid r^\star]\big]
 \;=\; -(1-\phi)\sum_{f\in F^\star}\operatorname{logit}(p_f).
\label{eq:bias}
\end{equation}
The bias is a genuine downward \emph{loss} for individually probable findings
($p_f>\tfrac12$) and grows as $\phi\to0$. A score that depends only on
observed-present evidence (explained-overlap, or PPR seeded on $F_\phi$) contains
no term indexed by $f\notin F_\phi$ and incurs no such penalty.
\end{proposition}
\noindent\textit{Proof in \Cref{app:proof}.} A correctly specified posterior that
\emph{marginalizes} unobserved findings carries no such term;
\Cref{ssec:qmr} confirms it empirically (the closed-world posterior collapses
under heavy under-observation while the marginalized and observation-aware
posteriors track the cheap positive-evidence priors, \Cref{fig:qmr}). The lesson
for buffered, partially observed diagnosis is to model the observation process,
not to treat unobserved evidence as negative evidence.

\section{Reconciliation as Structural Repair: Testing the Bridging Hypothesis}
\label{sec:bridging}

We apply the model of \Cref{sec:model} to the motivating PCAP instance.
Reconciliation is the dominant accuracy lever (\Cref{tab:ablation}), and
\Cref{hyp:perc} suggests a mechanism: it might \emph{bridge} fragmented cited
evidence into one component, raising the giant-component fraction $\Phimax$. We
test this on $200$ paired (pre/post-reconciliation) cases from two configuration
families ($\mathrm{E}\!\to\!\mathrm{F}$, $\mathrm{B}\!\to\!\mathrm{C}$, $100$ pairs
each; pre-reconciliation approximated by the no-reconcile sibling), building the
cited subgraph (\Cref{def:evgraph}) and its four order parameters and assessing
pre/post shifts (Wilcoxon) and structure--quality association (Spearman vs.\
weighted $\fone$).

\paragraph{Connectivity is unchanged; orphans are pruned.}
The giant-component fraction $\Phimax$ does \emph{not} shift across reconciliation
(Wilcoxon $p=0.67,0.71$); what moves is the orphan fraction $\iota$ (in
$\mathrm{E}\!\to\!\mathrm{F}$, $0.149\!\to\!0.069$, $z=-2.66$, $p=0.008$).
Reconciliation acts not as a bridge-builder but as a \emph{pruner of dangling
evidence}, refuting the strong bridging hypothesis.

\paragraph{Post-reconciliation structure predicts correctness.}
While the pre/post shift is small, the \emph{absolute} post-reconciliation
structure is strongly associated with quality: span density $\delta$ correlates
with $\fone$ at $\rho=0.64$--$0.67$, $\Phimax$ at $0.59$--$0.61$, and fragmentation
$\kappa$ negatively at $-0.58$--$-0.62$ (\Cref{fig:bridging}). A connected, dense
cited subgraph is thus a correlate of a correct diagnosis, consistent with
\Cref{hyp:perc}---but this is \emph{associational} on synthetic families: as
\Cref{sec:signals} shows, such associations are substantially driven by capture
size, so we do not claim connectivity as a size-independent predictor. The robust,
size-insensitive finding here is the \emph{mechanistic} one: reconciliation prunes
orphan evidence rather than bridging components.


\section{Structural Reliability Signals: A Controlled Negative Result}
\label{sec:signals}

It is tempting to read \Cref{hyp:perc} as a recipe for \emph{trust}: if a
correct diagnosis percolates, then connectivity or its stability should predict
correctness on real data, yielding a cheap, model-agnostic confidence signal.
From the model of \Cref{sec:model} we derive six candidate reliability signals
and ask, on the $103$-case golden set (base accuracy $0.767$), whether any of
them carries information about correctness \emph{beyond trivial capture size}.
The honest answer is no. Several signals correlate with correctness in-sample,
but once we control for capture size and elementary citation statistics, apply
capture-level permutation tests, compare against degree-preserving negative-control
nulls, and correct for multiple comparisons, \emph{none} survives. We report this
as a cautionary negative result: on this dataset, the apparent ``structure
predicts trust'' signal is an artifact of how much evidence a capture contains.

\subsection{In-sample associations (uncontrolled)}
\label{ssec:sig-pos}
At the case level several candidates correlate with correctness
(\Cref{tab:signals}): fragmentation ($\rho=+0.25$, $p=0.011$), the agreement of
two stochastic views of the cited evidence (edge-Jaccard $\rho=0.27$,
largest-component Jaccard $\rho=0.31$), and mean PageRank centrality
(\Cref{ssec:relevance}) of cited frames ($\rho=0.22$). Used as abstention features
they improve risk--coverage over the pipeline's own confidence
($\mathrm{AURC}=0.129$ vs $0.159$, \Cref{fig:riskcov}), and the gain replicates out
of sample (median held-out AURC $0.128$ vs $0.160$; the structural gate wins on
$97\%$ of $2{,}000$ splits). So there is a genuine, cheap triage signal. An earlier
version of this work stopped here and called these ``effective \emph{structural}
signals''. That label is wrong: \Cref{ssec:sig-neg,ssec:sig-integration} show the
gain is a capture-\emph{size} effect, not evidence-graph topology.

\subsection{The associations are a capture-size artifact}
\label{ssec:sig-neg}
The golden set is \emph{clustered}: each capture contributes several cases, and
captures vary widely in size. Larger captures cite more frames, are easier to
diagnose, and mechanically inflate every connectivity statistic, so an
uncontrolled case-level correlation conflates ``structure'' with ``amount of
evidence''. We therefore recompute each association as a partial Spearman
correlation controlling for capture size, citation count, and mean citation gap,
with significance assessed by a \emph{capture-level} permutation test (shuffling
correctness within captures) and Benjamini--Hochberg correction over all $14$
candidate constructions (\Cref{tab:controlled}).

The signal collapses. Fragmentation falls to $\rho=+0.14$ ($p=0.12$), centrality
to $\rho=+0.03$ ($p=0.35$), and connectivity-robustness to $\rho=+0.07$
($p=0.55$). Two degree-preserving \emph{negative controls} confirm the
diagnosis: the ``excess'' of $\Phimax$ over a rewired null carries no signal
($\rho=+0.04$, $p=0.73$), i.e.\ the raw connectivity correlation is entirely
explained by size and degree sequence. The two agreement metrics survive size
control at the raw level (largest-component Jaccard $\rho=+0.275$, $p=0.026$;
edge-Jaccard $\rho=+0.252$, $p=0.025$) but \emph{not} multiple-comparison
correction (BH $q=0.18$ for both). \emph{No candidate survives BH-FDR.} The
working abstention gate of \Cref{ssec:sig-pos} is therefore best read as a
capture-size/separability proxy (larger, cleaner captures are easier and the
fragmentation score tracks that) rather than as a measurement of evidence-graph
coherence.

\subsection{No incremental value, and a confirmatory test fails}
\label{ssec:sig-integration}
A grouped (leave-one-capture-out) logistic gate tells the same story: adding any
single signal to a size-plus-citation baseline yields no robust out-of-sample gain
in correct/incorrect AUROC, the largest nominal increment (largest-component
Jaccard, $\Delta\mathrm{AUROC}=0.15$) again failing FDR. Because the agreement
metrics were the only candidates with raw $p<0.05$ after size control, we
subjected them to a confirmatory test on the $n{=}76$ cases with a second-opinion
view. It fails both pre-specified checks. \textbf{Specificity:} once we control
for plain \emph{set-overlap} (node-Jaccard, a scalar the pipeline already records),
the \emph{structural} agreement loses significance ($\rho:0.28\!\to\!0.20$,
$p:0.025\!\to\!0.088$)---the graph framing adds nothing beyond ``do the two views
cite the same frames''. \textbf{Replication:} the incremental AUROC is positive in
only $77\%$ of $400$ held-out capture splits, below our $80\%$ bar. The test is
also badly underpowered (collinear metrics, $\mathrm{VIF}\approx3.4$; detecting the
observed residual $\rho\approx0.20$ would need $\sim\!650$ captures,
$\approx10\times$ this corpus). We therefore fold no structural term into the
confidence score and draw no positive claim.

\paragraph{Takeaways.} On this golden set, structural connectivity and its
perturbation stability provide \emph{no} reliability information beyond capture
size and trivial set-overlap; null-model normalization of $\Phimax$ adds nothing
(cited subgraphs sit near a connectivity ceiling), and motif surprise vanishes
under size control. The model's self-reported confidence is non-discriminative
(AUROC $0.49$, \Cref{fig:calib}); the deterministic composite orders better (AUROC
$0.60$) but is badly under-confident, useful for \emph{ranking}, not as a
probability. These negative results sharpen the thesis: the percolation prior
earns its keep in \emph{recoverability} (\Cref{sec:bench}) and \emph{localization}
(\Cref{sec:cross}), not as a drop-in trust score on an already-strong pipeline.


\section{A Synthetic Percolation Benchmark and the Detectability Threshold}
\label{sec:bench}

The PCAP study establishes correlation between connectivity and correctness but
cannot isolate the underlying phase transition, because the ground-truth latent
graph and the observation density $\phi$ are not controllable. We therefore
construct a synthetic benchmark in which both are known, allowing a direct test
of \Cref{hyp:perc} and \Cref{eq:phic}. Each \emph{incident graph} places a root
cause $R$ that seeds $L$ causal paths of intermediate symptoms toward an alarm
$A$, embedded in a background Erd\H{o}s--R\'enyi graph of mean degree $c$. Under
\Cref{ass:obs} each vertex is observed independently with probability $\phi$. We
then measure two distinct quantities: whether the incident is
\emph{recoverable} (an observed path connects $R$ to $A$), and whether a given
localizer actually \emph{identifies} $R$.

\paragraph{The empirical critical density approaches theory.}
The susceptibility (the mean size of the non-giant component containing a
random vertex, \Cref{sec:bg}) peaks at an empirical critical density
that approaches the theoretical value $\phi_c=1/c=0.25$ from above as the system
grows: $\hat\phi_c \approx 0.38, 0.36, 0.32, 0.28$ for
$N=200,500,1000,2000$ (\Cref{fig:pc}). This monotone approach is the expected
finite-size scaling behavior and confirms that the benchmark reproduces the
percolation transition predicted by \Cref{eq:phic}; via Molloy--Reed the same
critical point holds for any configuration-model degree distribution, so the
transition is generic rather than an Erd\H{o}s--R\'enyi artifact. Recoverability
crosses $50\%$ near $\phi\approx0.5$.

\paragraph{An algorithmic detectability gap above the percolation threshold.}
Recoverability is necessary but not sufficient for diagnosis: even when an
$R$--$A$ path is observed, a localizer must select $R$ from among many
candidates. To stress this distinction we introduce \emph{decoy hubs}, vertices
of high degree that are unrelated to the incident and clustered in a reserved
region of the graph, and compare localizers (\Cref{tab:detect},
\Cref{fig:detect}). The connectivity-aware personalized PageRank of
\Cref{eq:ppr} tracks the Bayesian oracle closely (peak top-1 accuracy $0.625$
vs.\ $0.675$ at $N=500$), whereas \emph{degree-only} heuristics collapse
(global degree $0.213$, component degree $0.20$) because they are captured by
the decoys; only PPR and the oracle ever exceed $50\%$ accuracy (finite
$p_{50}$). The benchmark thus isolates a second threshold beyond percolation: an
\emph{algorithmic} detectability gap separating relevance-weighted walks from
degree heuristics, which we observe again in every cross-domain experiment.


\section{Cross-Domain Evidence: the Principle Beyond Networking}
\label{sec:cross}

The motivating instance and the controlled benchmark both suggest that
relevance-weighted connectivity on a partially observed graph is a general
diagnostic primitive. We probe that by applying the \emph{same} model
(\Cref{def:evgraph}) and localizer (\Cref{eq:ppr}) to six external domains chosen
for topological diversity---directed service graphs, undirected infrastructure
networks, bipartite cause--finding graphs, program-dependence graphs,
retrieved-evidence graphs, and security attack graphs---without per-domain tuning
beyond edge orientation. The localization result is positive across all six;
whether structure also beats a size or separability baseline is domain-dependent
(\Cref{tab:crossdomain}). Microservice RCA and power grids use \emph{real} data;
software fault localization, multi-hop QA, and security attack-paths are
\emph{controlled simulations}---all evidence that the percolation \emph{principle}
is not networking-specific, not competitive entries against each field's tuned
state of the art. We develop the two domains closest to networking (microservice
RCA and bipartite QMR diagnosis) here and report the other four in
\Cref{app:crossdomain}; the synthesis below draws on all six.

\subsection{Microservice root-cause localization (Train Ticket chaos-injection)}
\label{ssec:openrca}
We use $451$ fault-injection incidents on the Train Ticket microservice
benchmark~\cite{zhou2021trainticket} ($64$ services), in which faults are
injected with Chaos~Mesh across $27$ resource-, network-, HTTP-, JVM-, and
DNS-level fault types.\footnote{This is a chaos-engineering corpus on Train
Ticket; it is \emph{not} the OpenRCA benchmark~\cite{xu2025openrca}, which
covers Bank/Market/Telecom cloud incidents. We keep the directory label
\texttt{openrca2} only for artifact reproducibility and claim no identity with
OpenRCA.} Each incident provides a directed service-call/propagation graph over
span/pod/container/service components, a set of observed alarm components, and
the ground-truth injected (root-cause) service (mean $5.1$ candidate services
after canonicalizing spans and containers to their parent service). Localizers
rank services; we report top-$k$ accuracy, MRR, and mean percentile
(\Cref{tab:openrca}).

The best localizer is a \emph{reverse}-PageRank prior that propagates mass
\emph{up} the causal edges toward upstream sources, attaining $89.1\%$ top-1 and a
$96.1$st mean percentile, far above forward PageRank ($12.6\%$) and random
($35.0\%$). One caveat keeps the headline honest: this reverse prior is
\emph{evidence-blind} (uniform restart, never reading the alarms), and seeding the
walk at the observed alarms attains only $53.0\%$. On this corpus it is edge
\emph{orientation}, not evidence-seeding, that carries the signal---the propagation
graphs are built around the injected root, so it sits near the structural source
and ranks highly under any reverse prior (\Cref{tab:openrca}). The result thus
confirms that orientation matters (the directed form of \Cref{eq:ppr-series}); a
cleaner demonstration of \emph{evidence-seeded} relevance appears in the RAG domain
(\Cref{ssec:rag}), where seeding is decisive.

\paragraph{Comparison to domain-native baselines (PetShop, RCAEval).} The Train
Ticket corpus ships graphs but no metric time series, so it cannot drive the
metric-based causal RCA methods that define the state of the art. On two public,
metric-rich benchmarks that can---PetShop~\cite{hardt2024petshop} and RCAEval RE2
Train Ticket~\cite{pham2025rcaeval}---our training-free relevance prior is a
\emph{competitive} zero-training localizer: it stays orientation-sensitive on
PetShop and matches the mid-tier causal baseline CIRCA on RCAEval (AC@1 $0.29$
vs.\ $0.32$, far above MicroCause/RCD), but it does \emph{not} beat a well-tuned
causal/statistical detector such as BARO ($0.67$). On metric-rich data a causal
model localizes \emph{which} node more accurately---the same
structure-vs-separability pattern as software fault localization
(\Cref{ssec:softfl}). \Cref{ssec:petshop} gives the full setup, per-method
tables, and the orientation analysis.

\subsection{Bipartite disease--symptom diagnosis (QMR): synthetic and real}
\label{ssec:qmr}
The third domain instantiates the bipartite causal setting of
\Cref{eq:ppr,eq:noisyor} and provides the most direct test of
\Cref{prop:bias}. We evaluate both a controlled generative model and a
templated real-world disease--symptom dataset.

\paragraph{Synthetic QMR.}
A generative noisy-OR model ($300$ diseases, $1000$ findings, $1000$ sampled
cases) permits comparison of a graph-only PPR relevance prior against the exact
noisy-OR posterior (\Cref{tab:qmr}). PPR attains $70.5\%$ top-1 accuracy (versus
$0.5\%$ for random selection) and tracks the posterior's ranking closely
($99.3$rd percentile). A simple \emph{explained-overlap} count (the number of
observed findings adjacent to a candidate disease) attains $94.4\%$, exceeding
PPR because high-degree ``hub'' findings dilute the relevance walk while leaving
the overlap count unaffected. This is the bipartite instance of the dual role
of high-degree nodes identified in \Cref{sec:bench}.

\paragraph{Real disease--symptom data.}
We repeat the evaluation on a $41$-disease, $131$-symptom, $4{,}920$-case
disease--symptom dataset~\cite{itachi2019dataset}. We caution that this is a
\emph{templated} symptom-signature dataset (not a clinical patient cohort) and
that the link weights $\hat\Prob[\text{symptom}\mid\text{disease}]$ are estimated
\emph{in-sample}, so the noisy-OR scores are optimistic upper bounds and the
informative quantity is the \emph{relative} robustness of the solvers under the
sparsity sweep below. With that caveat, graph-only PPR localizes the true disease
at $86.1\%$ top-1 ($33\times$ the $2.6\%$ chance rate) while raw degree is
uninformative ($2.4\%$): it is \emph{relevance-weighted} connectivity, not degree,
that carries the signal. The genuine hub symptoms (fatigue and vomiting each span
$17$ diseases) reproduce the hub-dilution effect, with PPR ($0.861$) just below
overlap ($0.972$).

\paragraph{Robustness under partial observation confirms \Cref{prop:bias}.}
Sweeping the fraction $\phi$ of symptoms reported (\Cref{fig:qmr}) isolates the
misspecification: the \emph{closed-world} noisy-OR collapses from $1.000$ at full
observation to $0.359$ at $\phi{=}0.2$ (it charges the true disease for
expected-but-unseen symptoms, \Cref{eq:noisyor}), whereas the correctly specified
posteriors of \Cref{rem:misspec} do not---the observation-aware model is the
\emph{most} robust ($0.728$) and the marginal model degrades gracefully
($0.628$). The cheap positive-evidence priors (overlap $0.653$, graph-only PPR
$0.603$) need \emph{no} knowledge of the observation process yet stay far above the
closed-world model. Since heavy under-observation is the defining regime of
buffered capture and sampled telemetry, the recipe is to model the observation
process or use a positive-evidence relevance score, never to treat unobserved
evidence as negative.



The QMR detectability collapse (\Cref{ssec:qmr}) is not a separate phenomenon:
specializing the percolation threshold to the bipartite (disease--finding)
configuration model gives a closed-form $\phi_c^{\mathrm{bip}}$ that a finite-size
susceptibility sweep recovers from above, tying the collapse to the same theory
as the one-mode benchmark (\Cref{ssec:bip-threshold}).

\section{Discussion and Conclusion}
\label{sec:disc}

\paragraph{A single structural principle, and where it fails.}
Across a controlled benchmark and six external domains, one principle recurs:
relevance-weighted connectivity on a partially observed evidence graph is an
inexpensive, label-free, model-agnostic signal for \emph{recoverability}
(\Cref{sec:bench}) and \emph{localization} (\Cref{sec:cross}). Equally important
is where it \emph{fails}: on the golden set it yields no reliability (trust)
signal beyond capture size (\Cref{sec:signals}), and in software fault
localization the same order parameter predicts \emph{whether} a fault is
recoverable yet a graph-only walk loses to a separability-driven spectrum score at
saying \emph{which} statement it is (\Cref{ssec:softfl}). Three threshold
extensions sharpen the positive half---the QMR collapse \emph{is} the bipartite
percolation threshold, which stays conservative under anomaly-biased observation
and bounds a measured localizable phase boundary
(\Cref{ssec:bip-threshold,ssec:mar,ssec:phaseboundary}).

\paragraph{The dual role of high-degree vertices.}
High-degree vertices aid localization when causal (cascading lines) and impede it
when incidental (decoy hubs, hub findings, high-fan-out spans). Degree alone is
thus an unreliable localizer; a walk that respects connectivity and, where
available, edge \emph{orientation} is the robust alternative---orientation alone
explains the reverse prior's $89\%$ top-1 on Train Ticket, while evidence-seeding
adds most where the graph is not already built around the cause (multi-hop RAG,
\Cref{ssec:rag}).

\paragraph{Closed-world fragility, and the value of a negative result.}
The most consequential theoretical result is \Cref{prop:bias} (confirmed in
\Cref{fig:qmr}): the closed-world noisy-OR posterior, optimal at full observation,
becomes \emph{less} robust than cheap positive-evidence priors when evidence is
missing (unobserved), not absent---a misspecification of the observation process,
not a failure of Bayes (\Cref{rem:misspec}). Our most carefully controlled finding
is itself negative: the intuitive ``structure predicts trust'' hypothesis does not
survive capture-size and set-overlap control on the golden set (\Cref{sec:signals}),
where every candidate fails capture-level permutation tests after
multiple-comparison correction---reported prominently because the opposite
in-sample reading is easy to reach and easy to publish.

\subsection{Limitations}
\label{ssec:limits}
The golden set is modest ($n\approx103$ cases, $87$ captures) and single-domain,
which bounds the negative result of \Cref{sec:signals} both ways: the
size-controlled signals add nothing detectable, but the study is underpowered for
a small residual effect (detecting the observed residual $\rho\approx0.20$ would
need $\sim\!650$ captures, $\approx\!10\times$ this corpus), so we claim only that
structure provides \emph{no demonstrable} reliability signal beyond capture size
here. Minor caveats---temporal/conversation-only PCAP edges, a no-reconcile sibling
proxy for pre-reconciliation graphs, templated disease signatures, and imperfect
external benchmarks---bound the remaining claims; none of the cross-domain results
is state-of-the-art, but together they evidence \emph{portability}: a single
untuned structural prior is broadly effective.

\paragraph{Conclusion.}
Diagnosis under partial observation is percolation on a partially observed
evidence graph: relevance-weighted connectivity is a portable signal for
\emph{recoverability} and \emph{localization}---but not, on our golden set, for
\emph{trust}. Across microservices, power grids, and real disease diagnosis the
same prior beats the closed-world noisy-OR posterior under under-observation
(\Cref{prop:bias}), where treating unobserved evidence as negative evidence is what
fails.


\bibliographystyle{tmlr}
\bibliography{references}

\appendix
\section{Deferred Derivations and Methods}
\label{app:methods}

This appendix collects the standard derivations, the formal definitions, and the
evaluation protocol underlying the main text; each is referenced where it is
used.

\subsection{Generating-function derivation of the percolation threshold}
\label{app:gf}
The modern, degree-distribution-agnostic derivation of \Cref{eq:phic} uses
probability generating functions~\cite{newman2001random,newman2003structure}.
Let $p_k$ be the probability that a uniformly chosen vertex has degree $k$, and
define the degree and excess-degree generating functions
\begin{equation}
G_0(x)=\sum_{k\ge 0}p_k x^k,
\qquad
G_1(x)=\frac{G_0'(x)}{G_0'(1)} = \frac{1}{\langle k\rangle}\sum_{k\ge 1}k\,p_k\,x^{k-1},
\label{eq:genfun}
\end{equation}
where $\langle k\rangle = G_0'(1)$. Following an edge to a vertex and counting
its \emph{other} edges yields the branching process governed by $G_1$; a giant
component exists iff its mean offspring exceeds one, $G_1'(1)>1$, the
Molloy--Reed criterion $\langle k^2\rangle - 2\langle k\rangle > 0$
~\cite{molloy1995critical}. Independently retaining each edge with probability
$\phi$ (bond percolation) rescales the branching ratio to $\phi\,G_1'(1)$, giving
the threshold $\phi_c=1/G_1'(1)$ of \Cref{eq:phic}~\cite{callaway2000network,
cohen2000resilience}. For a Poisson degree distribution of mean $c$,
$\langle k^2\rangle-\langle k\rangle = c^2$, hence $\phi_c=1/c$.

\subsection{Proof of \texorpdfstring{\Cref{prop:bias}}{Proposition 1}}
\label{app:proof}
Restricted to $F^\star$, full observation contributes $\sum_{f\in F^\star}\log p_f$;
under partial observation an observed finding still contributes $\log p_f$ while an
unobserved one, scored absent, contributes $\log(1-p_f)$. The per-finding
difference is $\log p_f-\log(1-p_f)=\operatorname{logit}(p_f)$, incurred with
probability $1-\phi$; truly-absent findings ($f\notin F^\star$) are scored absent
in both cases and cancel. Taking expectations over the independent observation
indicators gives \Cref{eq:bias}. Overlap and PPR are functions of $F_\phi$ alone,
so no absent-as-missing term appears. \qed

\subsection{The noisy-OR likelihood}
\label{app:noisyor}
In bipartite cause$\to$finding settings, let $q_{r,f}=\Prob[f\mid r]$ be the
per-link activation reliability and $\ell_f$ the per-finding leak (background)
probability. Under the noisy-OR assumption of independent causal activation, the
single-cause log-likelihood of a present/absent finding partition is
\begin{equation}
\log \Prob[F\mid r] = \underbrace{\sum_{f\in F}\log\!\big(1-(1-\ell_f)(1-q_{r,f})\big)}_{\text{present-finding support}}
 + \underbrace{\sum_{f\notin F}\log\!\big((1-\ell_f)(1-q_{r,f})\big)}_{\text{absent-finding penalty}},
\label{eq:noisyor}
\end{equation}
with posterior $\Prob[r\mid F]\propto \Prob[r]\,\Prob[F\mid r]$ and
$p_f := \Prob[f{=}1\mid r^\star]=1-(1-\ell_f)(1-q_{r^\star,f})$ in
\Cref{prop:bias}. A weakly activated finding ($p_f<\tfrac12$) scored as absent
actually \emph{raises} the likelihood, so the bias of \Cref{eq:bias} is not
unconditionally signed; for the individually-probable findings typical of QMR it
is a downward loss of order $\Theta((1-\phi)|F^\star|)$.

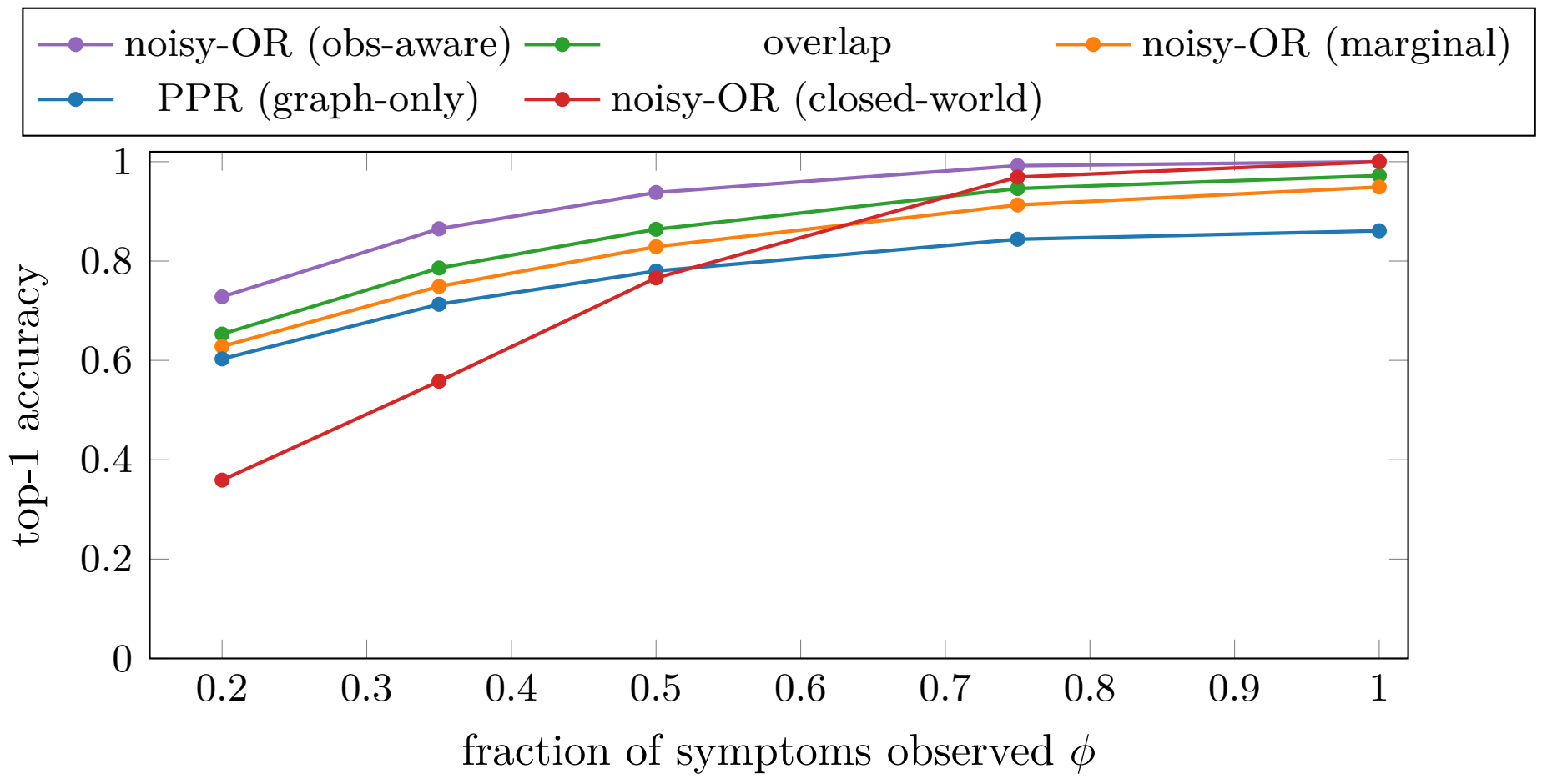
\begin{figure}[t]
\centering
\begin{tikzpicture}
\begin{axis}[
  width=0.7\textwidth,height=5.6cm,
  xlabel={fraction of symptoms observed $\phi$}, ylabel={top-1 accuracy},
  xmin=0.15,xmax=1.02, ymin=0,ymax=1.02,
  legend style={at={(0.5,1.03)},anchor=south,font=\footnotesize,legend columns=3},
]
\addplot[diaglines,c3] coordinates {(1.0,1.000)(0.75,0.992)(0.5,0.938)(0.35,0.865)(0.2,0.728)};
\addplot[diaglines,c2] coordinates {(1.0,0.972)(0.75,0.946)(0.5,0.864)(0.35,0.786)(0.2,0.653)};
\addplot[diaglines,c4] coordinates {(1.0,0.949)(0.75,0.913)(0.5,0.829)(0.35,0.749)(0.2,0.628)};
\addplot[diaglines,c0] coordinates {(1.0,0.861)(0.75,0.844)(0.5,0.780)(0.35,0.713)(0.2,0.603)};
\addplot[diaglines,c1] coordinates {(1.0,1.000)(0.75,0.969)(0.5,0.766)(0.35,0.558)(0.2,0.359)};
\legend{noisy-OR (obs-aware), overlap, noisy-OR (marginal), PPR (graph-only), noisy-OR (closed-world)}
\end{axis}
\end{tikzpicture}
\caption{Real disease--symptom data under under-observation (\Cref{ssec:qmr}).
The \emph{closed-world} noisy-OR (red) collapses as symptoms are dropped (the
misspecification of \Cref{prop:bias}) while the correctly specified
\emph{observation-aware} posterior (top) degrades most gracefully and the
\emph{marginal} posterior and the cheap positive-evidence priors (overlap, PPR)
sit in between. The fragility is in the closed-world reading, not in Bayesian
inference.}
\label{fig:qmr}
\end{figure}

\begin{remark}[Misspecification vs.\ a failure of Bayes]
\label{rem:misspec}
\Cref{eq:bias} is a property of the \emph{closed-world} reading, not of Bayesian
inference. A correctly specified posterior that marginalizes unobserved findings
scores only observed-present evidence and never charges $r^\star$ for an
expected-but-unseen finding (it omits the $\log(1-p_f)$ terms). \Cref{ssec:qmr}
confirms this: under heavy under-observation the closed-world posterior collapses,
whereas the marginalized and observation-aware posteriors degrade gracefully,
tracking the positive-evidence priors (\Cref{fig:qmr}).
\end{remark}

\subsection{Configuration-model nulls and motif surprise}
\label{ssec:null}
To ask whether an observed connectivity value exceeds its degree sequence, we
compare against a \emph{configuration-model} null that fixes the degree sequence
but rewires edges uniformly at random~\cite{molloy1995critical,newman2003structure}.
For a statistic $x$ with null mean $\mu_0$ and standard deviation $\sigma_0$
(estimated over rewired replicas), we report the standardized surprise
$z=(x-\mu_0)/\sigma_0$ and its empirical percentile. \emph{Motif surprise}
applies this to ordered subtype bigrams $g=(a\!\to\!b)$ over the observation
timeline, scoring the configuration-null $z$-score of each bigram count; high $z$
flags anomalous repetition (e.g.\ runs of \texttt{qos\_data}). Both are evaluated
as candidate reliability signals in \Cref{sec:signals}.

\subsection{Evaluation protocol and metrics}
\label{ssec:metrics}
For a localizer producing a ranking of candidate causes, \emph{top-$k$ accuracy}
is the fraction of cases whose true cause appears in the top $k$; \emph{mean
reciprocal rank} (MRR) is $\frac1N\sum_i 1/\mathrm{rank}_i$; and \emph{mean
percentile} is the average of $1-(\mathrm{rank}_i-1)/(m_i-1)$ over cases with
$m_i$ candidates. Monotone association between a continuous signal and
correctness is measured by the Spearman rank correlation $\rho$ with its
two-sided $p$-value; paired pre/post shifts are tested with the Wilcoxon
signed-rank test. For reliability we adopt selective
prediction~\cite{elyaniv2010foundations,geifman2017selective}: cases are admitted
in descending score order and the \emph{risk--coverage} curve plots error rate
against coverage $\tau\in(0,1]$, summarized by the area under it
$\mathrm{AURC}=\int_0^1 \mathrm{risk}(\tau)\,d\tau$ (lower is better) and by
accuracy at fixed coverage. Triage of correct vs.\ incorrect cases is measured by
the AUROC of the signal. All randomized procedures are seeded; correlations are
computed after removing undefined (NaN) entries.

\section{Additional Cross-Domain Studies and Theory Extensions}
\label{app:crossdomain}

This appendix reports in full the four cross-domain studies summarized in \Cref{tab:crossdomain}---power grids, spectrum-based software fault localization, multi-hop retrieval-augmented QA, and security attack-path attribution---together with three threshold extensions (the bipartite cause--finding threshold, biased non-MCAR observation, and the detectability phase boundary). They are placed here so that the main text can stay focused on the networking instance and its two closest analogues (microservice RCA and bipartite QMR diagnosis), without discarding the evidence that the percolation principle is not networking-specific.

\subsection{A bipartite percolation threshold for cause--finding diagnosis}
\label{ssec:bip-threshold}
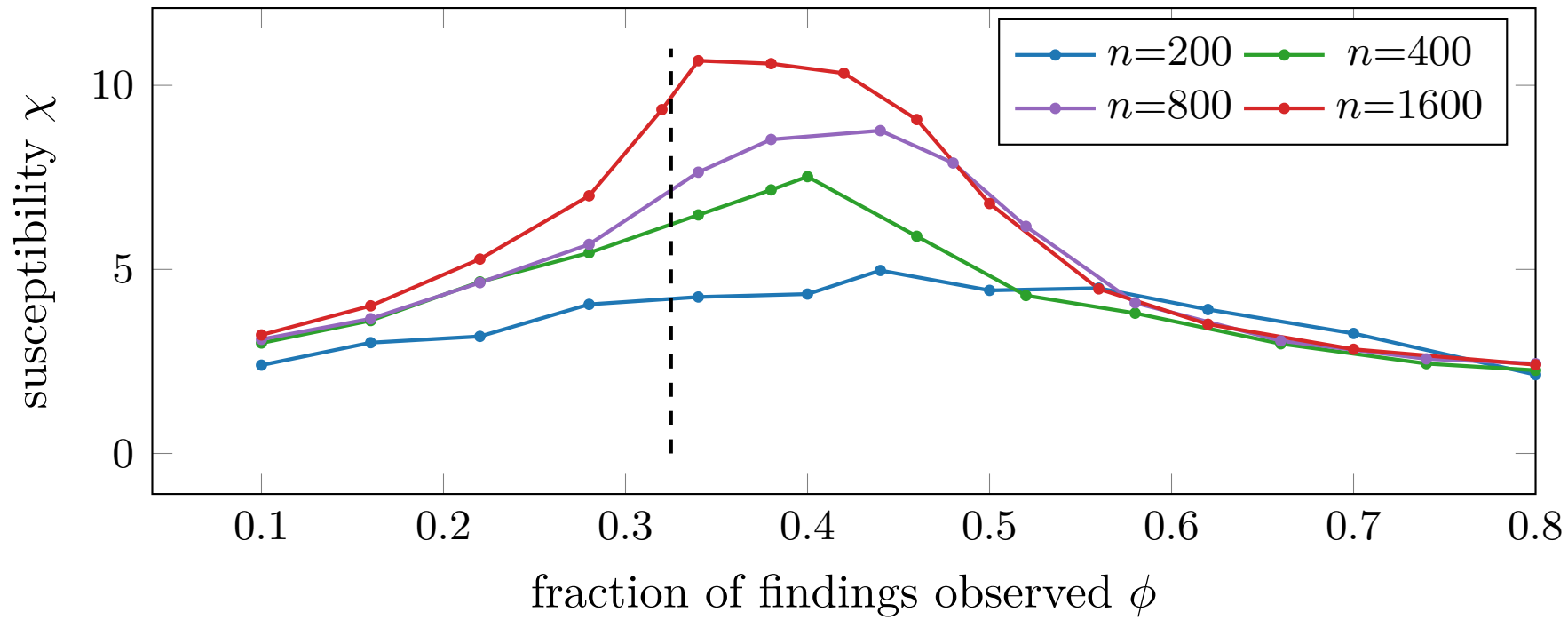
\begin{figure}[t]
\centering
\begin{tikzpicture}
\begin{axis}[width=0.72\textwidth,height=5.2cm,
  xlabel={fraction of findings observed $\phi$},
  ylabel={susceptibility $\chi$}, xmin=0.04,xmax=0.8,
  legend style={at={(0.98,0.98)},anchor=north east,font=\footnotesize,legend columns=2}]
\addplot[diaglines,c0,mark size=0.8pt] coordinates {(0.10,2.40)(0.16,3.01)(0.22,3.18)(0.28,4.05)(0.34,4.25)(0.40,4.33)(0.44,4.97)(0.50,4.43)(0.56,4.49)(0.62,3.91)(0.70,3.26)(0.80,2.14)};
\addplot[diaglines,c2,mark size=0.8pt] coordinates {(0.10,3.00)(0.16,3.61)(0.22,4.66)(0.28,5.45)(0.34,6.48)(0.38,7.16)(0.40,7.52)(0.46,5.90)(0.52,4.29)(0.58,3.81)(0.66,2.98)(0.74,2.44)(0.80,2.26)};
\addplot[diaglines,c3,mark size=0.8pt] coordinates {(0.10,3.10)(0.16,3.66)(0.22,4.64)(0.28,5.68)(0.34,7.64)(0.38,8.53)(0.44,8.77)(0.48,7.89)(0.52,6.17)(0.58,4.09)(0.66,3.06)(0.74,2.57)(0.80,2.44)};
\addplot[diaglines,c1,mark size=0.8pt] coordinates {(0.10,3.22)(0.16,4.01)(0.22,5.28)(0.28,7.00)(0.32,9.34)(0.34,10.67)(0.38,10.59)(0.42,10.33)(0.46,9.07)(0.50,6.79)(0.56,4.47)(0.62,3.51)(0.70,2.83)(0.80,2.41)};
\addplot[dashed,black,thick,forget plot] coordinates {(0.325,0)(0.325,11)};
\legend{$n{=}200$,$n{=}400$,$n{=}800$,$n{=}1600$}
\end{axis}
\end{tikzpicture}
\caption{Bipartite percolation (\Cref{ssec:bip-threshold}): the susceptibility
peak sharpens and moves toward the predicted $\phi_c^{\mathrm{bip}}=0.325$
(dashed) as $n$ grows.}
\label{fig:bipthr}
\end{figure}

The QMR detectability sweep (\Cref{ssec:qmr}) is governed by the same percolation
theory, specialized to the bipartite (disease--finding) configuration model. Let
the disease and finding excess-degree branching factors be
$f_1'(1)=\langle j(j{-}1)\rangle/\langle j\rangle$ and
$g_1'(1)=\langle k(k{-}1)\rangle/\langle k\rangle$. A giant component exists iff
$f_1'(1)\,g_1'(1)>1$~\cite{newman2001random}; if each \emph{finding} is observed
independently with probability $\phi$ (finding-side site percolation, exactly the
partial-observation model of \Cref{ssec:qmr}), the disease-to-disease branching
ratio rescales to $\phi\,f_1'(1)g_1'(1)$, giving the bipartite threshold
\begin{equation}
\phi_c^{\mathrm{bip}}=\frac{1}{f_1'(1)\,g_1'(1)}.
\label{eq:phicbip}
\end{equation}
Simulating finding-side occupation on the disease one-mode projection, the
finite-size susceptibility peak converges to \Cref{eq:phicbip} from above
(\Cref{fig:bipthr}; tabulated in \Cref{tab:bipthr}), the same data-collapse logic
that locates $\phi_c=1/c$ on the one-mode benchmark, now tying the QMR collapse to
the theory.

\subsection{Comparison to domain-native baselines (PetShop and RCAEval)}
\label{ssec:petshop}
The Train Ticket corpus of \Cref{ssec:openrca} ships propagation graphs but no
metric time series, so it cannot drive the metric-based causal RCA methods that
constitute the state of the art. To compare our relevance prior against
\emph{domain-native} baselines on a level field, we use the public PetShop
benchmark~\cite{hardt2024petshop}: $68$ injected performance issues on a
$41$-component microservice application, each with a service-call graph,
normal/abnormal CloudWatch metrics, and a ground-truth root-cause \emph{node}.
We evaluate inside PetShop's own harness (node-level top-$k$ recall), detecting
anomalies with a robust per-metric $z$-score and ranking candidates by a
personalized-PageRank walk seeded on the anomalous nodes (\Cref{tab:petshop});
we also report a degree heuristic, an anomaly-only ranking that drops the graph,
and the walk run in reverse.

The results line up with the rest of the paper. The relevance prior transfers and
stays orientation-sensitive: the forward walk, mass flowing along call edges
toward dependencies, gives the best top-1 of the structure-only solvers ($0.31$)
and far exceeds the reverse walk ($0.06$), because PetShop's root cause is a
\emph{downstream} dependency rather than the upstream source of the chaos corpus.
But structure does not beat domain-native causal RCA: the published
graph-traversal and CIRCA baselines exceed $0.8$ top-3 recall on most
scenarios~\cite{hardt2024petshop}, well above our walk's $0.44$. The bottleneck is
diagnostic, not topological: an anomaly-only ranking caps at $0.41$ top-3, so the
true root cause is often \emph{not} locally anomalous (it manifests downstream),
and a positive-evidence walk cannot implicate a quiet-but-causal node the way a
counterfactual or interventional model can. This is the microservice form of the
broader thesis: structure says \emph{whether} and roughly \emph{where} a fault is
recoverable, but on metric-rich data a causal model says \emph{which} node more
accurately (cf.\ the SBFL result of \Cref{ssec:softfl}). We make no
state-of-the-art claim; the value is a faithful, same-harness comparison that
bounds what structure alone buys---and since our front-end is a plain $z$-score
rather than PetShop's SCM-based scorers, it isolates the \emph{ranking} stage, not
the anomaly detector.

\paragraph{Same-system check (RCAEval RE2 Train Ticket).} PetShop is a different
application; for a same-system comparison we use the public RCAEval RE2 Train
Ticket benchmark~\cite{pham2025rcaeval} ($90$ cases, six fault types), which is
metric-rich and so supports the causal/statistical baselines that define the state
of the art. Building the call graph from trace spans (parent-span $\to$ child-span
service), seeding the walk on $z$-score-flagged services, and reporting
service-level AC@$k$ (RCAEval's own metric, \Cref{tab:rcaeval}), our training-free
prior attains AC@1 $=0.29$: it \emph{matches} CIRCA ($0.32$) and far exceeds
MicroCause ($0.10$) and RCD ($0.09$), though it trails the strongest statistical
baseline, the BARO change-point detector ($0.67$)~\cite{pham2024baro}. Two details
sharpen the PetShop reading. Degree alone collapses to AC@1 $=0.04$. Without
evidence, topology is nearly useless here, exactly as the null-model analysis of
\Cref{sec:bench} predicts. And unlike PetShop, the graph adds little \emph{over} a
plain anomaly-only ranking ($0.29$ vs.\ $0.30$), because RE2's resource faults make
the injected service its own strongest anomaly, leaving the walk little to recover.
The thread across both benchmarks is the same: an evidence-seeded structural prior
is a competitive, zero-training localizer on par with mid-tier causal baselines,
but it neither replaces nor beats a well-tuned causal detector on metric-rich
data.

\subsection{Power-grid cascading failures (PowerGraph)}
\label{ssec:power}
\begin{figure}[t]
\centering
\begin{tikzpicture}
\begin{axis}[
  width=0.7\textwidth,height=5.6cm,
  xlabel={fraction of lines removed}, ylabel={giant-component fraction},
  xmin=0,xmax=0.6, ymin=0,ymax=1.02,
  legend style={at={(0.98,0.98)},anchor=north east},
]
\addplot[diaglines,c1] coordinates {
 (0,1)(0.05,0.992)(0.10,0.983)(0.15,0.941)(0.20,0.924)(0.25,0.907)
 (0.30,0.602)(0.35,0.356)(0.40,0.271)(0.45,0.271)(0.50,0.212)(0.55,0.144)(0.60,0.110)};
\addplot[diaglines,c0] coordinates {
 (0,1)(0.05,0.994)(0.10,0.977)(0.15,0.964)(0.20,0.929)(0.25,0.877)
 (0.30,0.813)(0.35,0.693)(0.40,0.570)(0.45,0.457)(0.50,0.321)(0.55,0.284)(0.60,0.196)};
\legend{targeted (betweenness), random}
\end{axis}
\end{tikzpicture}
\caption{IEEE-118 robustness: targeted removal of high-betweenness lines
fragments the grid far faster than random removal.}
\label{fig:power}
\end{figure}
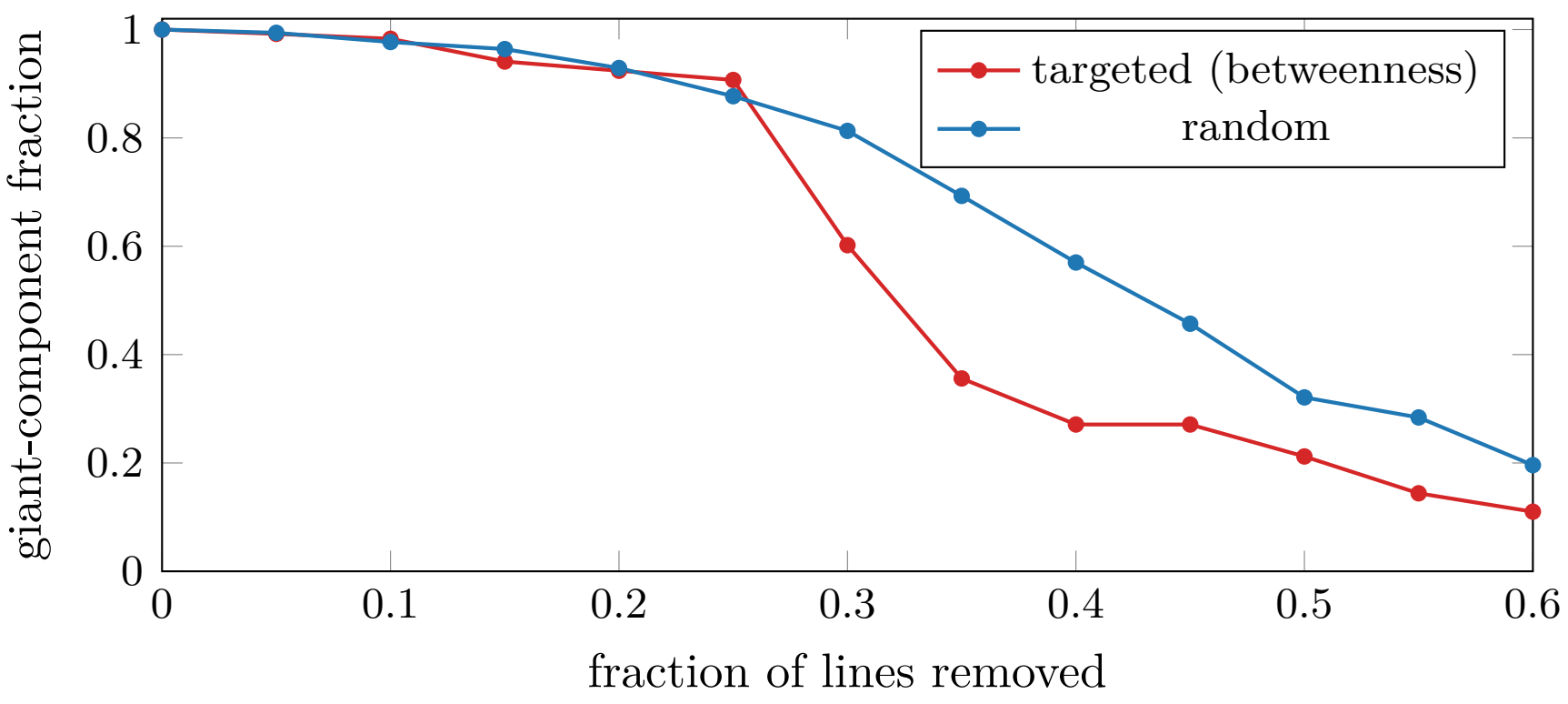

We analyze four grids (IEEE-39, IEEE-118, IEEE-24, and the UK network) from
PowerGraph~\cite{varbella2024powergraph}, treating each as an undirected
bus--branch graph annotated with cascading-edge labels and demand-not-served
(DNS) severity. Here the structural primitive is edge betweenness
(\Cref{sec:bg}) rather than a relevance walk, and we test three predictions
that the percolation viewpoint makes about infrastructure (\Cref{tab:power},
\Cref{fig:power}):
\begin{enumerate}[leftmargin=1.6em,itemsep=2pt]
\item[\textbf{B1}] \textbf{Central lines participate in cascades.} Edge
betweenness ranks the lines that appear in cascades with AUC $0.713$ (IEEE-118),
$0.699$ (IEEE-39), and $0.622$ (IEEE-24); only the dense, mesh-like UK grid is
near chance ($0.586$), consistent with betweenness losing discriminative power
when many redundant paths exist.
\item[\textbf{B2}] \textbf{Severity is only weakly structural.} The betweenness
of the \emph{initially} tripped line predicts DNS severity only weakly
(Spearman $-0.098$ to $+0.293$): the centrality of the trigger does not
determine the magnitude of the eventual cascade, which depends on load and
protection dynamics beyond pure topology.
\item[\textbf{B3}] \textbf{Targeted removal dominates random removal.} Deleting
lines in descending betweenness order fragments the giant component far faster
than random deletion---on IEEE-118 the giant-component fraction falls to $0.356$
at $35\%$ targeted removal versus $0.693$ under random removal---reproducing the
robustness asymmetry of \Cref{eq:phic}~\cite{albert2000error,callaway2000network,
cohen2000resilience} on real infrastructure.
\end{enumerate}

\begin{table}[t]
\centering
\caption{Power-grid cascades (PowerGraph). B1: AUC of edge betweenness ranking
cascading lines. B2: Spearman of tripped-line betweenness vs.\ DNS severity.}
\label{tab:power}
\small
\begin{tabular}{lccccc}
\toprule
Grid & nodes & edges & cascade events & B1 AUC & B2 $\rho$ \\
\midrule
IEEE-118 & 118 & 186 & 1167 & \textbf{0.713} & $-0.098$ \\
IEEE-39  & 39  & 46  & 281  & 0.699 & $+0.139$ \\
IEEE-24  & 24  & 38  & 3444 & 0.622 & $+0.205$ \\
UK       & 29  & 99  & 4531 & 0.586 & $+0.293$ \\
\bottomrule
\end{tabular}
\end{table}

\subsection{Software fault localization (spectrum-based)}
\label{ssec:softfl}
Our fourth domain is spectrum-based fault localization (SBFL), where the
``evidence graph'' is a program-dependence graph and the evidence is the set of
statements exercised by failing tests. We run a controlled simulation (in the
spirit of the synthetic QMR study, not the real Defects4J~\cite{just2014defects4j}
JVM harness): $2880$ programs over $72$ configurations spanning program size,
suite size, average dependence degree, fault-trigger probability, and a
flaky-failure rate that scatters the failing evidence. Each program has one
planted fault; localizers rank statements and we report the standard SBFL metrics
(top-$N$, MRR, and the EXAM wasted-effort fraction). We compare the classic
alarm-blind spectrum formulas---Tarantula~\cite{jones2005tarantula},
Ochiai~\cite{abreu2007ochiai}, and D${}^\star$~\cite{wong2014dstar}---against the
same relevance walk used throughout (PPR over the dependence graph, seeded by the
failing spectrum) and against Ochiai with a structural tie-break.

This domain reproduces \emph{both} halves of our thesis. First, the positive
half: the percolation order parameter of the \emph{suspicious set} (the top-$q$
statements by Ochiai, induced on the dependence graph) predicts whether the fault
is localizable. When the fault sits in the giant component of the suspicious set,
Ochiai top-1 accuracy is $0.49$ ($n{=}1720$); when the suspicious set fragments
and the fault falls into a smaller piece, accuracy collapses to $0.03$
($n{=}164$). The fault's component fraction correlates with localizability at
Spearman $\rho{=}{+}0.28$ (95\% CI $[0.24,0.32]$) against top-1 and
${+}0.45$ $[0.41,0.49]$ against $1-\textrm{EXAM}$---the same
``evidence percolates $\Rightarrow$ recoverable'' diagnostic the paper derives
for diagnosis, now in a fourth, unrelated domain.

Second, the negative half, mirroring \Cref{sec:signals}: structure does
\emph{not} beat a simple separability baseline. A graph-only relevance walk is
clearly \emph{worse} than the native spectrum formulas (\Cref{tab:softfl}), and
adding a structural tie-break to Ochiai buys only ${+}0.01$ top-1 and nothing on
EXAM. The single strongest predictor of localizability is not topological at all:
it is the size of the fault's spectrum tie-group---a separability/size quantity,
$\rho{=}{+}0.56$ $[0.53,0.60]$---just as capture size dominated every structural
trust signal on the golden set. Connectivity again says \emph{whether} a fault
can be found, but a domain-native score, not graph structure, says \emph{which
statement it is}.

\begin{table}[t]
\centering
\caption{Spectrum-based fault localization ($2880$ simulated programs). The
graph relevance walk (PPR) underperforms the domain's native spectrum formulas;
a structural tie-break on Ochiai adds at most $0.01$ top-1. Structure predicts
recoverability (text) but does not beat the separability-driven formulas at
pinpointing the fault. EXAM is wasted effort (lower is better).}
\label{tab:softfl}
\small
\begin{tabular}{l cccc c}
\toprule
Solver & top-1 & top-3 & top-5 & MRR & EXAM\\
\midrule
random                     & 0.014 & 0.042 & 0.076 & 0.069 & 0.497\\
coverage ($e_f$)           & 0.092 & 0.331 & 0.436 & 0.248 & 0.265\\
PPR (graph, evidence-seed) & 0.100 & 0.294 & 0.406 & 0.237 & 0.298\\
PPR (graph, Ochiai-seed)   & 0.202 & 0.393 & 0.465 & 0.327 & 0.274\\
Tarantula                  & 0.184 & 0.402 & 0.501 & 0.328 & 0.241\\
D${}^\star$ ($\star{=}2$)  & 0.271 & 0.441 & 0.490 & 0.381 & 0.251\\
Ochiai                     & 0.292 & 0.480 & 0.530 & 0.409 & 0.238\\
\textbf{Ochiai + struct.\ tie-break} & \textbf{0.302} & 0.475 & 0.526 & \textbf{0.411} & 0.255\\
\bottomrule
\end{tabular}
\end{table}

\subsection{Multi-hop retrieval-augmented QA (evidence graphs)}
\label{ssec:rag}
Our fifth domain maps the thesis onto retrieval-augmented generation, where
multi-hop questions (in the style of HotpotQA~\cite{yang2018hotpotqa} /
2WikiMultiHop~\cite{ho2020twowiki}) are answerable
only if the retrieved evidence \emph{connects} the question entity to the answer
through a chain of supporting facts. We simulate this directly (a knowledge graph,
a planted reasoning chain, a noisy retriever with imperfect recall and hard
entity-sharing distractors, and a probabilistic reader whose accuracy depends on
chain completeness and distractor load---deliberately \emph{not} on graph
connectivity, so percolation is a genuine predictor rather than a definition):
$3240$ questions over $108$ configurations, reader base accuracy $0.60$.

This is the domain where structure is \emph{most} valuable, on both claims.
\textbf{(B)} The percolation order parameter of the retrieved-evidence graph
sharply predicts correctness: when the question and answer entities fall in the
same retrieved component, reader accuracy is $0.88$ ($n{=}1992$); when the
evidence fragments, $0.14$ ($n{=}1248$). Question--answer connectivity correlates
with correctness at Spearman $\rho{=}{+}0.73$ (95\% CI $[0.70,0.76]$) and
$\Phi_{\mathrm{gcc}}$ at ${+}0.49$ $[0.44,0.53]$, whereas the size baseline
(inverse number of retrieved facts) is \emph{negatively} associated
(${-}0.16$). Unlike the golden set and SBFL, here a percolation signal beats a
size baseline outright. \textbf{(A)} A relevance walk seeded at the question
entities recovers the gold supporting facts far better than the lexical-overlap
baseline that production RAG uses (\Cref{tab:rag}): recall@$L$ rises from $0.35$
(lexical) to $0.52$ (PPR), and MRR from $0.63$ to $0.82$.

\begin{table}[t]
\centering
\caption{Multi-hop RAG ($3240$ simulated questions). Structure-aware selection
(degree and the relevance walk) recovers the gold supporting facts well above the
lexical baseline used by production retrievers. recall@$L$ = fraction of the $L$
gold facts ranked in the top $L$; mean-rank-frac is lower-is-better.}
\label{tab:rag}
\small
\begin{tabular}{l ccc}
\toprule
Solver & recall@$L$ & MRR & mean-rank-frac\\
\midrule
random            & 0.213 & 0.417 & 0.560\\
lexical overlap   & 0.347 & 0.634 & 0.471\\
degree            & 0.594 & 0.853 & 0.292\\
\textbf{PPR (question-seeded)} & \textbf{0.516} & \textbf{0.818} & \textbf{0.304}\\
PPR (lexical-seeded)           & 0.518 & 0.819 & 0.304\\
\bottomrule
\end{tabular}
\end{table}

\subsection{Security alert correlation and attack-path attribution}
\label{ssec:secpath}
Our sixth domain is intrusion response: an attacker moves laterally along an
attack graph~\cite{sheyner2002attackgraphs} from an initial compromise
(``patient zero'') to a target; sensors
raise alerts on some compromised nodes (imperfect detection) plus false positives
(noise), and the defender must name the root compromise. We simulate $4860$
incidents over $162$ configurations. The localizer is the same relevance walk,
run \emph{against} the attack edges (toward sources), mirroring the
reverse-PageRank result on microservice RCA (\Cref{ssec:openrca}).

This domain reproduces the dual pattern most cleanly. \textbf{Orientation is
decisive:} the reverse walk attains a $0.92$ mean percentile and $0.185$ top-1,
versus $0.039$ top-1 for the forward walk (\Cref{tab:secpath})---a
$4.7\times$ gap, exactly as on the Train Ticket corpus. \textbf{Percolation predicts confident
attribution:} $\Phi_{\mathrm{gcc}}$ of the alert-evidence graph correlates with
reverse-walk success at $\rho{=}{+}0.42$ $[0.39,0.45]$. \textbf{But structure does
not beat a separability baseline:} the strongest single predictor is again a
size/separability quantity---fewer correlated alerts, $\rho{=}{+}0.48$
$[0.45,0.51]$---and a plain alert-degree heuristic matches the reverse walk on
top-1 ($0.195$ vs $0.185$). As on the golden set and in SBFL, connectivity says
\emph{whether} attribution is recoverable, but a domain-native separability score
is as good at pinpointing the root.

\begin{table}[t]
\centering
\caption{Security attack-path attribution ($4860$ simulated incidents). The
reverse relevance walk (against attack edges) dominates the forward walk and
attains the best mean percentile, but does not beat a plain alert-degree
heuristic at top-1. Naming patient-zero exactly is hard; percentile localization
is strong.}
\label{tab:secpath}
\small
\begin{tabular}{l cccc}
\toprule
Solver & top-1 & top-3 & MRR & mean pctile\\
\midrule
random        & 0.009 & 0.028 & 0.049 & 0.498\\
alert-degree  & \textbf{0.195} & 0.461 & 0.365 & 0.834\\
PPR (forward) & 0.039 & 0.294 & 0.223 & 0.817\\
\textbf{PPR (reverse)} & 0.185 & 0.423 & 0.359 & \textbf{0.922}\\
\bottomrule
\end{tabular}
\end{table}

\subsection{Threshold under biased (non-MCAR) observation}
\label{ssec:mar}
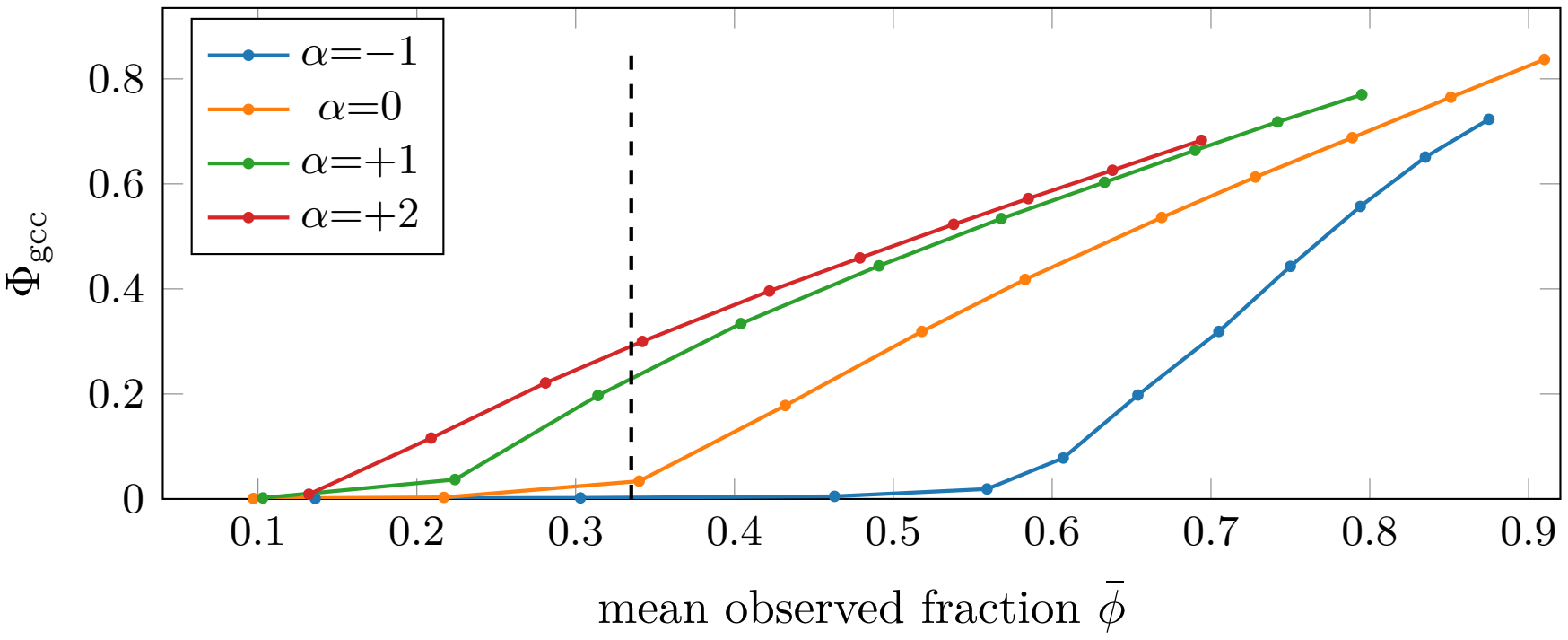
\begin{figure}[t]
\centering
\begin{tikzpicture}
\begin{axis}[width=0.72\textwidth,height=5.2cm,
  xlabel={mean observed fraction $\bar\phi$},
  ylabel={$\Phi_{\mathrm{gcc}}$}, xmin=0.04,xmax=0.92,ymin=0,
  legend style={at={(0.02,0.98)},anchor=north west,font=\footnotesize}]
\addplot[diaglines,c0,mark size=0.8pt] coordinates {(0.136,0.001)(0.303,0.002)(0.463,0.005)(0.559,0.019)(0.607,0.078)(0.654,0.198)(0.705,0.319)(0.750,0.443)(0.794,0.557)(0.835,0.651)(0.875,0.723)};
\addplot[diaglines,c4,mark size=0.8pt] coordinates {(0.097,0.001)(0.217,0.003)(0.340,0.034)(0.432,0.178)(0.518,0.319)(0.583,0.418)(0.669,0.536)(0.728,0.613)(0.789,0.688)(0.851,0.765)(0.910,0.837)};
\addplot[diaglines,c2,mark size=0.8pt] coordinates {(0.103,0.002)(0.224,0.037)(0.314,0.197)(0.404,0.334)(0.491,0.444)(0.568,0.534)(0.633,0.603)(0.690,0.664)(0.742,0.718)(0.795,0.770)};
\addplot[diaglines,c1,mark size=0.8pt] coordinates {(0.132,0.009)(0.209,0.116)(0.281,0.221)(0.342,0.300)(0.422,0.396)(0.479,0.459)(0.538,0.523)(0.585,0.572)(0.638,0.626)(0.694,0.683)};
\addplot[dashed,black,thick,forget plot] coordinates {(0.335,0)(0.335,0.85)};
\legend{$\alpha{=}{-}1$,$\alpha{=}0$,$\alpha{=}{+}1$,$\alpha{=}{+}2$}
\end{axis}
\end{tikzpicture}
\caption{Biased observation shifts the giant-component onset: hub-biased
sampling ($\alpha{>}0$) percolates left of the MCAR threshold (dashed),
anti-correlated sampling ($\alpha{=}{-}1$) to its right.}
\label{fig:mar}
\end{figure}

Assumption~\ref{ass:obs} (MCAR) is wrong in the motivating domain: a buffer
captures \emph{around} an anomaly, over-representing evidence near the relevant
subgraph and near hub vertices. We re-derive the threshold for degree-correlated
site occupation. Occupying a degree-$k$ vertex with probability $\phi_k$, the
occupied cluster branches supercritically iff
\begin{equation}
R=\sum_k \frac{k\,p_k}{\langle k\rangle}\,\phi_k\,(k-1)>1,
\label{eq:Rmar}
\end{equation}
which reduces to $\phi\,G_1'(1)$ when $\phi_k\equiv\phi$. Parameterizing a bias
$\phi_k\propto(k/\langle k\rangle)^{\alpha}$ and solving \Cref{eq:Rmar} for the
critical \emph{mean} observed fraction $\bar\phi_c$, hub-biased observation
($\alpha>0$, the realistic buffer case) percolates---becomes
diagnosable---at a substantially \emph{lower} mean observation than MCAR predicts,
while anti-correlated observation is harder (\Cref{tab:mar}, \Cref{fig:mar}). The
prediction matches the simulated susceptibility peak in every case. This is both
an honest limitation (MCAR is conservative) and a usable correction: when the
observation process favors the anomaly neighborhood, the effective detectability
threshold drops by $40$--$60\%$.

\begin{table}[t]
\centering
\caption{Threshold under degree-biased observation ($\langle k\rangle{=}3$, MCAR
$\phi_c{=}0.335$). $\alpha{>}0$ over-samples hubs (realistic); the re-derived
$\bar\phi_c$ matches the simulated peak and falls well below MCAR.}
\label{tab:mar}
\small
\begin{tabular}{r ccc}
\toprule
$\alpha$ & $\bar\phi_c$ (theory) & $\bar\phi^\ast$ (sim) & vs MCAR\\
\midrule
$-1$ & 0.581 & 0.612 & $+73\%$ (harder)\\
$0$  & 0.335 & 0.372 & MCAR\\
$+1$ & 0.204 & 0.223 & $-39\%$ (easier)\\
$+2$ & 0.145 & 0.170 & $-57\%$ (easier)\\
\bottomrule
\end{tabular}
\end{table}

\subsection{The detectability gap as a phase boundary}
\label{ssec:phaseboundary}
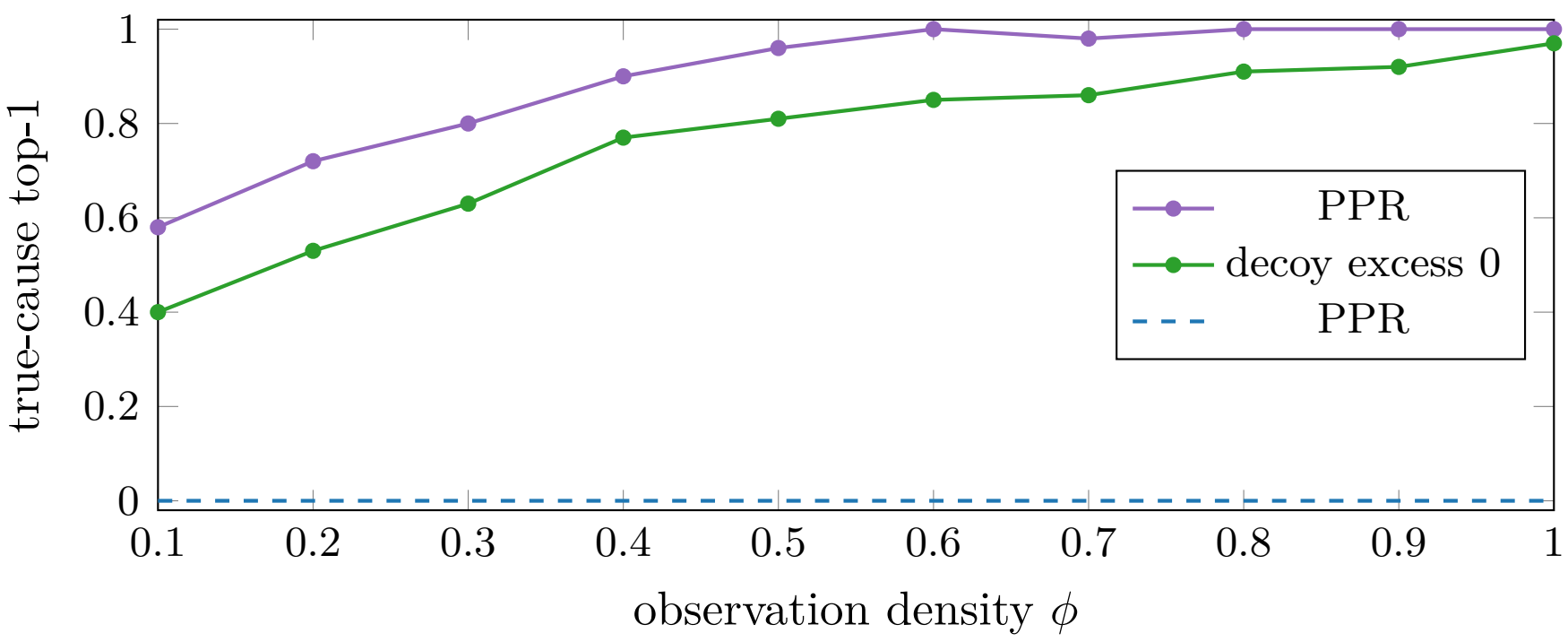
\begin{figure}[t]
\centering
\begin{tikzpicture}
\begin{axis}[width=0.72\textwidth,height=5.2cm,
  xlabel={observation density $\phi$}, ylabel={true-cause top-1},
  xmin=0.1,xmax=1.0, ymin=-0.02,ymax=1.02,
  legend style={at={(0.98,0.5)},anchor=east,font=\footnotesize}]
\addplot[diaglines,c3] coordinates {(0.1,0.58)(0.2,0.72)(0.3,0.80)(0.4,0.90)(0.5,0.96)(0.6,1.00)(0.7,0.98)(0.8,1.00)(0.9,1.00)(1.0,1.00)};
\addplot[diaglines,c2] coordinates {(0.1,0.40)(0.2,0.53)(0.3,0.63)(0.4,0.77)(0.5,0.81)(0.6,0.85)(0.7,0.86)(0.8,0.91)(0.9,0.92)(1.0,0.97)};
\addplot[dashed,c0,thick] coordinates {(0.1,0)(1.0,0)};
\legend{PPR, decoy excess $0$, PPR, decoy excess $16$, degree (any excess)}
\end{axis}
\end{tikzpicture}
\caption{Detectability gap: the relevance walk localizes the true cause above an
observation threshold that rises with decoy confusability, whereas the
evidence-blind degree heuristic is fooled by the decoy hub everywhere
(top-1 $\equiv 0$).}
\label{fig:phaseb}
\end{figure}

\Cref{sec:bench} reported, at one operating point, that a relevance walk localizes
the true cause where a degree heuristic is fooled by an incidental decoy hub. We
map this as a two-dimensional boundary over the decoy's excess degree (structural
confusability) and the observation density $\phi$. Across the entire plane the
evidence-blind degree heuristic stays in the undetectable phase (top-1
$\equiv 0$: the decoy hub always out-ranks the true cause), while the relevance
walk has a localizable phase whose high-confidence boundary (top-1 $\ge 0.8$)
\emph{rises} with decoy excess: $\phi^\ast \approx 0.4$ at zero excess, climbing
to $\approx 0.5$--$0.6$ as the decoy grows (\Cref{fig:phaseb}). This is the
``two thresholds'' picture made quantitative---an information/recoverability
threshold in $\phi$, and an algorithmic threshold that the relevance walk meets
but a centrality heuristic never does in the presence of a decoy hub.

\section{Supplementary Figures}
\label{app:figs}

These figures support the main-text results where they are discussed and are
collected here to keep the body compact; each is referenced at the point it is
used.

\begin{figure}[t]
\centering
\begin{tikzpicture}
\begin{axis}[
  width=0.6\textwidth, height=5.8cm,
  ybar=10pt, bar width=11pt,
  ymin=-0.85, ymax=0.85,
  ylabel={Spearman $\rho$ vs weighted $\fone$},
  xtick={1,2,3,4},
  xticklabels={$\Phimax$,$\kappa$,$\iota$,$\delta$},
  enlarge x limits=0.15,
  legend style={at={(0.5,1.02)},anchor=south,legend columns=2,font=\footnotesize},
  nodes near coords, nodes near coords style={font=\tiny},
  every node near coord/.append style={/pgf/number format/fixed,
     /pgf/number format/precision=2},
]
\addplot[fill=c0] coordinates {(1,0.589) (2,-0.579) (3,-0.412) (4,0.637)};
\addplot[fill=c4] coordinates {(1,0.615) (2,-0.618) (3,-0.349) (4,0.674)};
\legend{E$\rightarrow$F,B$\rightarrow$C}
\end{axis}
\end{tikzpicture}
\caption{Post-reconciliation connectivity predicts correctness
(\Cref{sec:bridging}). Span density $\delta$ and giant-component fraction
$\Phimax$ correlate positively with weighted $\fone$; fragmentation $\kappa$
correlates negatively. All $|\rho|\ge0.35$ are significant at $p<10^{-3}$.}
\label{fig:bridging}
\end{figure}
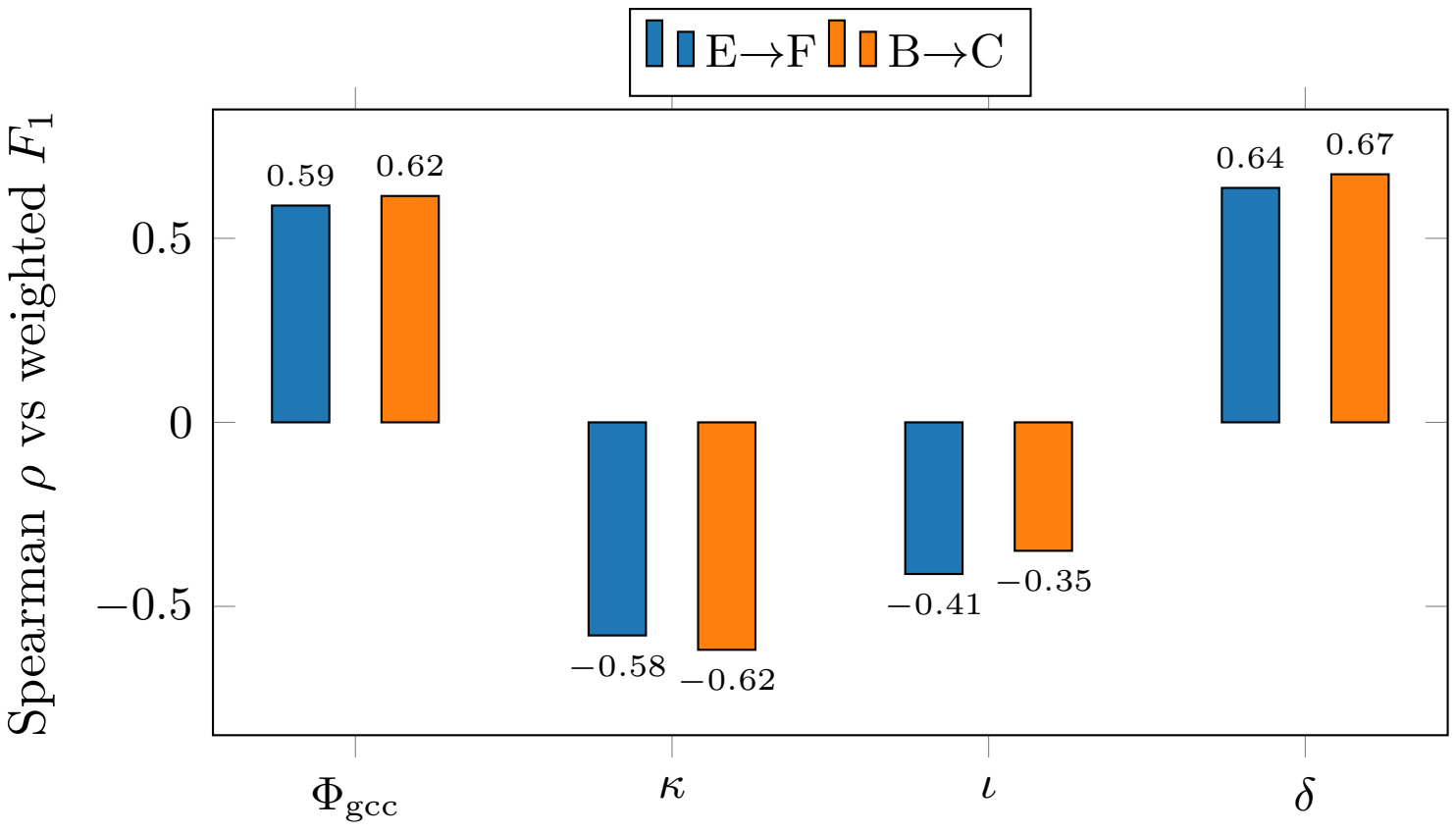

\begin{figure}[t]
\centering
\begin{subfigure}[t]{0.49\textwidth}
\centering
\begin{tikzpicture}
\begin{axis}[
  width=\textwidth, height=5.2cm,
  xlabel={self-reported confidence bucket},
  ylabel={accuracy},
  ymin=0,ymax=1.05, xmin=0.5,xmax=6.5,
  xtick={1,2,3,4,5,6},
  legend style={at={(0.98,0.02)},anchor=south east,font=\footnotesize},
]
\addplot[diaglines,c1] coordinates {(1,0.5)(2,0.4)(3,0.8)(4,1.0)(5,1.0)(6,1.0)};
\addplot[diaglines,c0] coordinates {(1,0.7)(2,0.7)(3,0.7)(4,0.7)(5,0.9)(6,1.0)};
\legend{self-reported, deterministic composite}
\end{axis}
\end{tikzpicture}
\caption{Calibration: self-reported confidence is non-monotone and
non-discriminative (AUROC $0.49$), while the deterministic composite is monotone
and orders better (AUROC $0.60$). Note the composite is a better \emph{ranker}
than \emph{probability}: it is systematically under-confident
(ECE $0.47$ vs $0.19$), so it should drive selective prediction, not be read as a
calibrated probability.}
\label{fig:calib}
\end{subfigure}\hfill
\begin{subfigure}[t]{0.49\textwidth}
\centering
\begin{tikzpicture}
\begin{axis}[
  width=\textwidth, height=5.2cm,
  xlabel={coverage},
  ylabel={accuracy on covered cases},
  ymin=0.74,ymax=0.9, xmin=0.45,xmax=1.02,
  legend style={at={(0.02,0.98)},anchor=north west,font=\footnotesize},
  x dir=reverse,
]
\addplot[diaglines,c0] coordinates {(1.0,0.767)(0.9,0.785)(0.8,0.780)(0.7,0.792)(0.6,0.790)(0.5,0.808)};
\addplot[diaglines,c2] coordinates {(1.0,0.767)(0.9,0.774)(0.8,0.768)(0.7,0.778)(0.6,0.823)(0.5,0.865)};
\addplot[diaglines,c3] coordinates {(1.0,0.767)(0.9,0.763)(0.8,0.805)(0.7,0.819)(0.6,0.871)(0.5,0.865)};
\legend{composite, $\kappa$-frac, rank-mean}
\end{axis}
\end{tikzpicture}
\caption{Risk--coverage: the structural abstention gate beats the pipeline's
confidence score at low coverage, and the gain \emph{replicates out of sample}
(median held-out AURC $0.13$ vs $0.16$; beats the composite in $97\%$ of capture
splits). \Cref{tab:controlled} shows this reflects capture size/separability, not
evidence-graph topology.}
\label{fig:riskcov}
\end{subfigure}
\caption{Reliability and selective prediction on the golden set
(\Cref{sec:signals}). The calibration motivation (left) stands; the
selective-prediction gain (right) is real and replicates out-of-sample, but
reflects capture size rather than evidence-graph topology (\Cref{tab:controlled}).}
\end{figure}
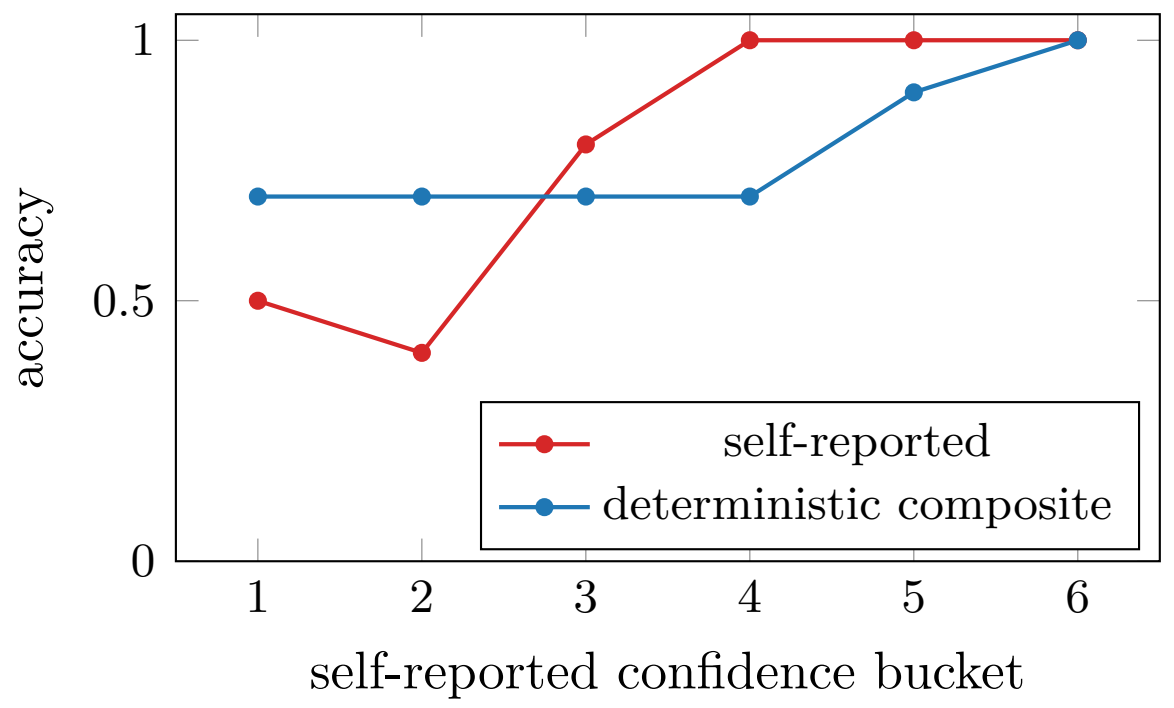
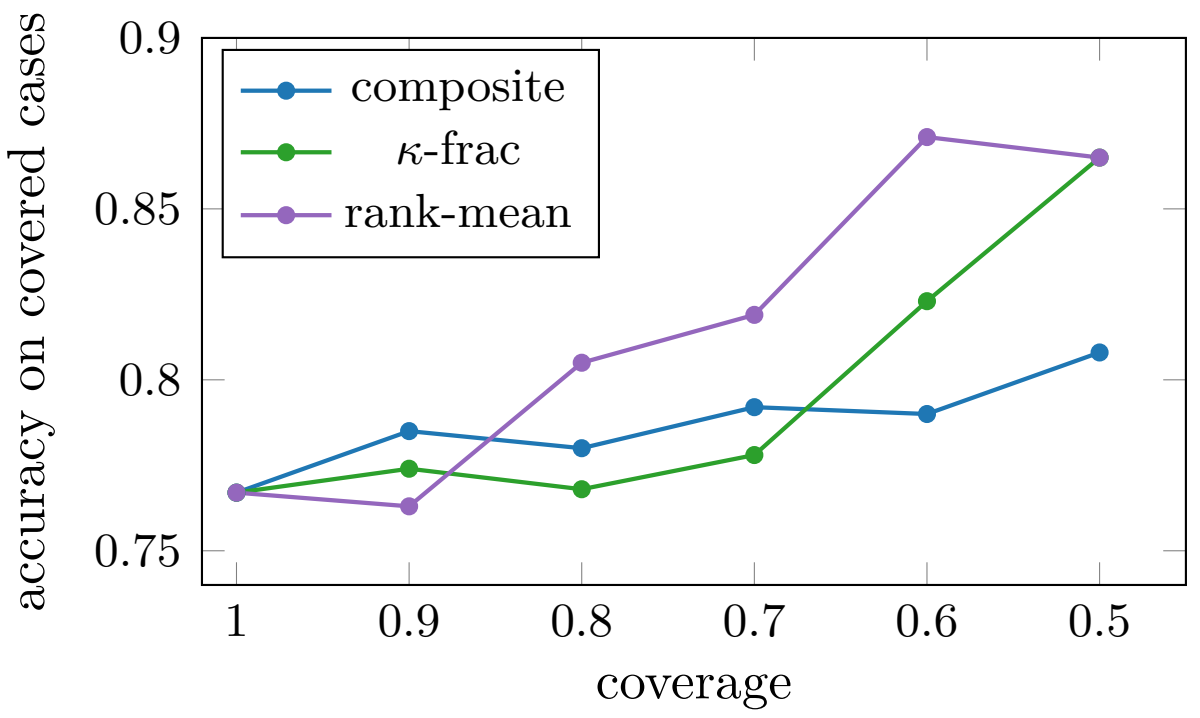

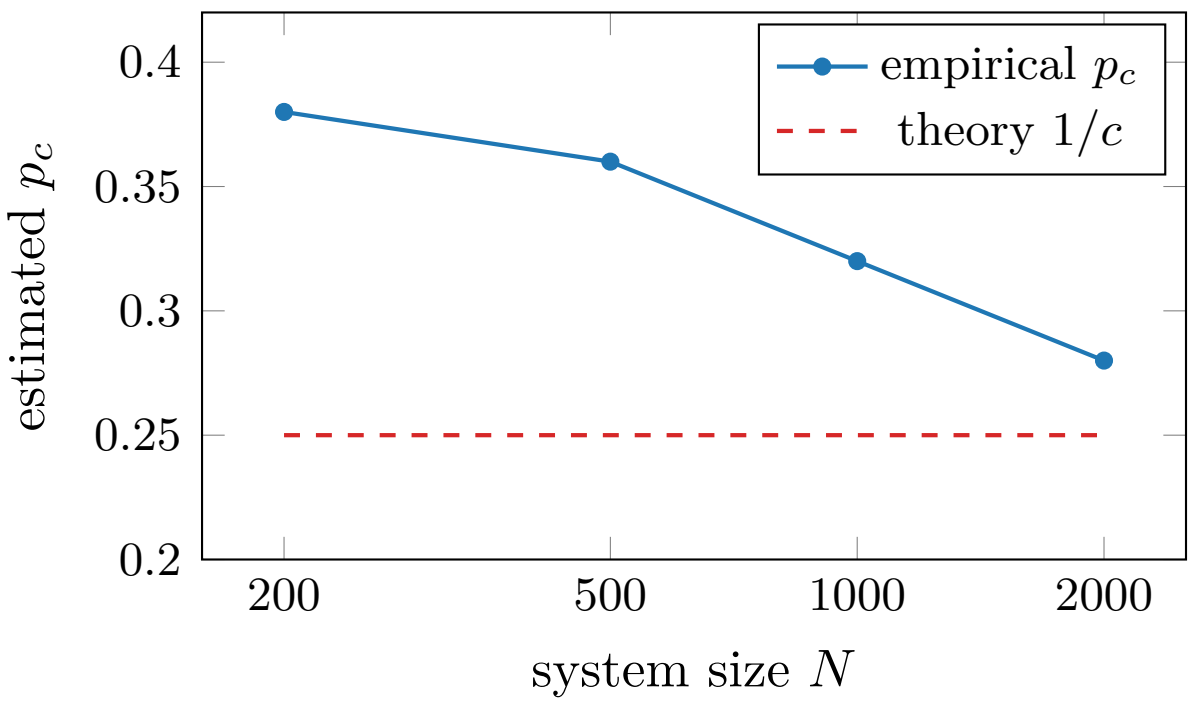
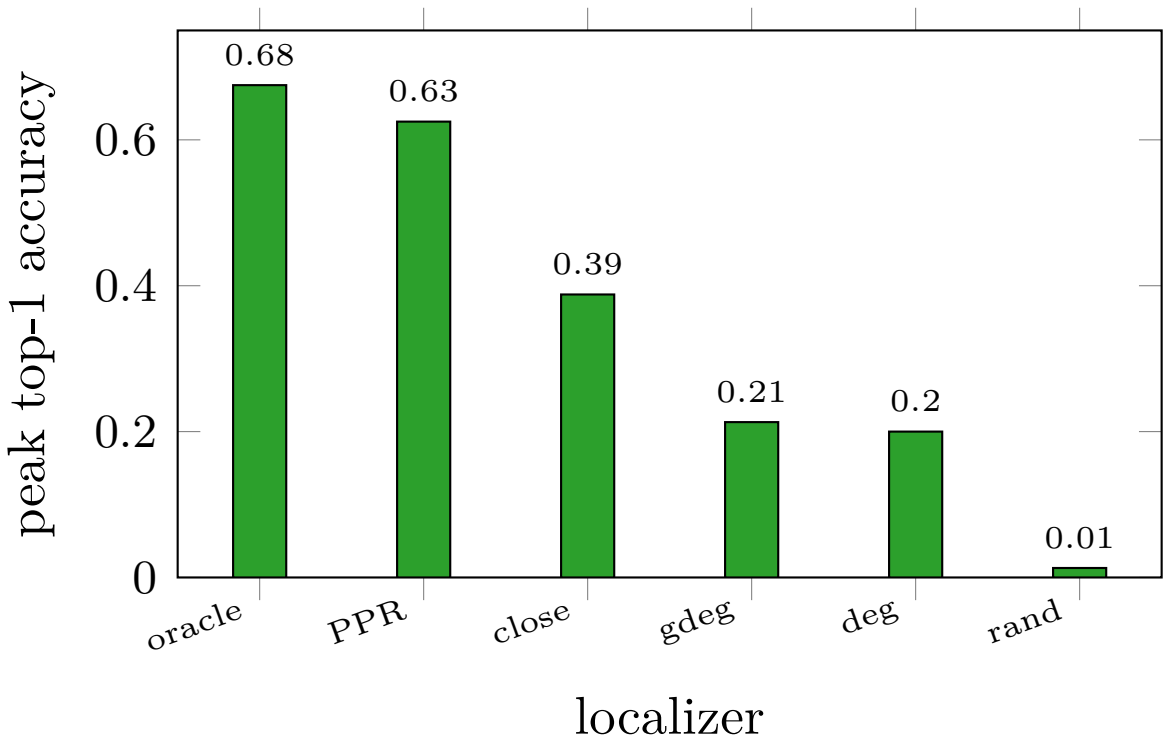
\begin{figure}[t]
\centering
\begin{subfigure}[t]{0.49\textwidth}
\centering
\begin{tikzpicture}
\begin{axis}[
  width=\textwidth,height=5.2cm,
  xlabel={system size $N$}, ylabel={estimated $p_c$},
  xmode=log, log basis x=2,
  xtick={200,500,1000,2000}, xticklabels={200,500,1000,2000},
  ymin=0.2,ymax=0.42,
  legend style={at={(0.98,0.98)},anchor=north east},
]
\addplot[diaglines,c0] coordinates {(200,0.38)(500,0.36)(1000,0.32)(2000,0.28)};
\addplot[dashed,c1,thick] coordinates {(200,0.25)(2000,0.25)};
\legend{empirical $p_c$, theory $1/c$}
\end{axis}
\end{tikzpicture}
\caption{Empirical critical density approaches $1/c=0.25$.}
\label{fig:pc}
\end{subfigure}\hfill
\begin{subfigure}[t]{0.49\textwidth}
\centering
\begin{tikzpicture}
\begin{axis}[
  width=\textwidth,height=5.2cm,
  ybar, bar width=10pt,
  xlabel={localizer}, ylabel={peak top-1 accuracy},
  symbolic x coords={oracle,PPR,close,gdeg,deg,rand},
  xtick=data, x tick label style={font=\tiny,rotate=20,anchor=east},
  ymin=0,ymax=0.75,
  nodes near coords, nodes near coords style={font=\tiny},
  every node near coord/.append style={/pgf/number format/fixed,/pgf/number format/precision=2},
]
\addplot[fill=c2] coordinates {(oracle,0.675)(PPR,0.625)(close,0.388)(gdeg,0.213)(deg,0.20)(rand,0.013)};
\end{axis}
\end{tikzpicture}
\caption{PPR tracks oracle; degree scheme fail under decoys.}
\label{fig:detect}
\end{subfigure}
\caption{Synthetic percolation benchmark (\Cref{sec:bench}): the empirical
critical density approaches $1/c$ (left), and a connectivity-aware localizer
tracks the Bayesian oracle while degree heuristics collapse under decoy hubs
(right).}
\end{figure}

\section{Supplementary Tables}
\label{app:tables}

These tables report the full numerical detail behind the figures and prose of the main text; each is referenced where it is discussed. They are collected here to keep the body focused on the headline results.

\begin{table}[t]
\centering
\caption{Cross-domain summary (\Cref{sec:cross}). The same evidence-graph model
and relevance walk, applied with only a per-domain choice of edge orientation,
localize the root cause well above chance in every domain; structure beats a
size/separability baseline only where the evidence is genuinely fragmented
(multi-hop RAG). The Train Ticket RCA and QMR studies are detailed in the main
text (\Cref{ssec:openrca,ssec:qmr}); the baseline comparison and the remaining
studies are reported in full in \Cref{app:crossdomain}.}
\label{tab:crossdomain}
\small
\begin{tabular}{@{}llll@{}}
\toprule
Domain & Data & Headline localization & Beats size baseline?\\
\midrule
Microservice RCA (Train Ticket) & real & reverse PPR, $0.89$ top-1 & orientation-driven\\
PetShop / RCAEval RE2 & real & evidence PPR matches CIRCA & no (causal RCA wins)\\
QMR disease--symptom & real+sim & evidence PPR, $0.86$ top-1 & overlap $\approx$ PPR\\
Power grid (PowerGraph) & real & edge betweenness, AUC $0.71$ & targeted $\gg$ random\\
Software fault loc.\ (SBFL) & sim & predicts recoverability & no (spectrum wins)\\
Multi-hop RAG & sim & PPR recall $0.35{\to}0.52$ & \textbf{yes}\\
Security attack-path & sim & reverse PPR, $0.92$ pctile & no (degree ties)\\
\bottomrule
\end{tabular}
\end{table}

\begin{table}[t]
\centering
\caption{Ablation of the evidence-grounded ensemble pipeline (from the
companion study \emph{PROBE}~\cite{henry2026probe}; weighted evidence $\fone$ is the
headline metric, \Cref{ssec:pcap}). Configurations differ in the number of drafts
and whether a reconciliation pass is applied. Reconciliation is the dominant
lever.}
\label{tab:ablation}
\small
\begin{tabular}{lcccccc}
\toprule
Config & $N$ & Key $\fone$ & Reliab.\ $\fone$ & \textbf{Wt $\fone$} & Perfect-key \% & role \\
\midrule
A & 104 & 0.773 & 0.460 & 0.871 & 45.2 & lean baseline \\
B & 104 & 0.680 & 0.831 & 0.912 & 34.6 & ensemble, no reconcile \\
\textbf{C} & 100 & 0.815 & 0.851 & \textbf{0.964} & 51.0 & ensemble + reconcile \\
D & 104 & 0.444 & 0.692 & 0.842 & 21.2 & ablation \\
E & 104 & 0.461 & 0.692 & 0.845 & 20.2 & ablation \\
F & 100 & 0.776 & 0.846 & 0.957 & 50.0 & ensemble + reconcile \\
\bottomrule
\end{tabular}
\end{table}

\begin{table}[t]
\centering
\caption{Detectability on the synthetic benchmark with decoy hubs
($N=500$, $c=4$, $80$ trials). Peak top-1 accuracy and the observation density
$p_{50}$ at which accuracy first reaches $50\%$ (``--'' $=$ never).}
\label{tab:detect}
\small
\begin{tabular}{lcc}
\toprule
Localizer & peak acc & $p_{50}$ \\
\midrule
oracle (Bayesian) & \textbf{0.675} & 0.482 \\
PageRank (connectivity-aware) & 0.625 & 0.493 \\
closeness & 0.388 & -- \\
global degree & 0.213 & -- \\
degree (component) & 0.200 & -- \\
random & 0.013 & -- \\
\bottomrule
\end{tabular}
\end{table}

\begin{table}[t]
\centering
\caption{Bipartite QMR diagnosis (full observation, \Cref{ssec:qmr}). A
graph-only relevance prior (PPR) localizes the true cause far above chance and
tracks the noisy-OR posteriors, on both a synthetic model and a real
disease--symptom dataset. At full observation the closed-world and
observation-aware posteriors coincide; the \emph{marginal} posterior is slightly
lower because it ignores true negatives. The closed-world/correctly-specified
split matters only under partial observation (\Cref{fig:qmr}). Real-data noisy-OR
scores are in-sample upper bounds.}
\label{tab:qmr}
\small
\begin{tabular}{l cccc c cccc}
\toprule
& \multicolumn{4}{c}{Synthetic ($300\times1000$)} & & \multicolumn{4}{c}{Real ($41\times131$, $4920$ cases)}\\
\cmidrule(lr){2-5}\cmidrule(lr){7-10}
Solver & top-1 & top-3 & MRR & pctile & & top-1 & top-3 & MRR & pctile\\
\midrule
random              & 0.002 & 0.005 & 0.018 & 0.505 & & 0.026 & 0.078 & 0.108 & 0.502\\
degree              & 0.001 & 0.001 & 0.036 & 0.547 & & 0.024 & 0.073 & 0.105 & 0.500\\
overlap             & 0.952 & 0.974 & 0.964 & 0.997 & & 0.972 & 1.000 & 0.986 & 0.999\\
\textbf{PPR (graph-only)} & \textbf{0.705} & 0.871 & 0.798 & 0.993 & & \textbf{0.861} & 0.888 & 0.899 & 0.990\\
noisy-OR (closed-world)   & 0.986 & 0.993 & 0.990 & 1.000 & & 1.000 & 1.000 & 1.000 & 1.000\\
noisy-OR (marginal)       & 0.973 & 0.988 & 0.982 & 0.999 & & 0.949 & 1.000 & 0.973 & 0.999\\
noisy-OR (obs-aware)      & 0.986 & 0.993 & 0.990 & 1.000 & & 1.000 & 1.000 & 1.000 & 1.000\\
\bottomrule
\end{tabular}
\end{table}

\begin{table}[t]
\centering
\caption{Public RCAEval RE2 Train Ticket benchmark~\cite{pham2025rcaeval}
($90$ cases), coarse-grained service-level AC@$k$. Our training-free relevance
prior matches CIRCA and beats MicroCause/RCD but trails the BARO change-point
detector; degree alone (no evidence) collapses. Baseline AC@1 values are as
reported by \citet{pham2025rcaeval}.}
\label{tab:rcaeval}
\small
\begin{tabular}{lcc}
\toprule
Method & AC@1 & AC@3 \\
\midrule
degree (structural only)        & 0.04 & 0.20 \\
anomaly-only (no graph)         & 0.30 & 0.79 \\
relevance walk, reverse         & 0.30 & 0.52 \\
\textbf{relevance walk (ours)}  & \textbf{0.29} & 0.61 \\
\midrule
\multicolumn{3}{l}{\emph{published baselines~\cite{pham2025rcaeval} (AC@1):}}\\
BARO~\cite{pham2024baro}        & 0.67 & -- \\
CIRCA                           & 0.32 & -- \\
MicroCause                      & 0.10 & -- \\
RCD                             & 0.09 & -- \\
\bottomrule
\end{tabular}
\end{table}

\begin{table}[t]
\centering
\caption{Bipartite threshold $\phi_c^{\mathrm{bip}}=1/(f_1'g_1')$ vs the
finite-size susceptibility-peak location (\Cref{ssec:bip-threshold}). The
empirical peak approaches the predicted $\phi_c^{\mathrm{bip}}\approx0.33$ from
above as the system grows.}
\label{tab:bipthr}
\small
\begin{tabular}{r cc}
\toprule
$n_{\mathrm{dis}}$ & $\phi_c^{\mathrm{bip}}$ (theory) & $\phi^\ast$ (susceptibility peak)\\
\midrule
200  & 0.364 & 0.524\\
400  & 0.296 & 0.395\\
800  & 0.337 & 0.434\\
1600 & 0.325 & \textbf{0.343}\\
\bottomrule
\end{tabular}
\end{table}

\begin{table}[t]
\centering
\caption{Bridging analysis on $200$ reconciliation pairs. $\Phimax$ does not
shift (Wilcoxon n.s.); the orphan fraction $\iota$ drops significantly in
E$\to$F. Post-reconciliation structure correlates strongly with weighted
$\fone$ (last column).}
\label{tab:bridging}
\small
\begin{tabular}{llcccc}
\toprule
Family & Param & mean pre & mean post & Wilcoxon $p$ & Spearman $\rho(\Phi_{\text{post}},\fone)$ \\
\midrule
\multirow{4}{*}{E$\to$F} & $\Phimax$ & 0.700 & 0.717 & 0.67 & $+0.589$ \\
 & $\kappa$ & 2.84 & 2.49 & 0.056 & $-0.579$ \\
 & $\iota$ & 0.149 & \textbf{0.069} & \textbf{0.008} & $-0.412$ \\
 & $\delta$ & 0.611 & 0.644 & 0.27 & $+0.637$ \\
\midrule
\multirow{4}{*}{B$\to$C} & $\Phimax$ & 0.694 & 0.681 & 0.71 & $+0.615$ \\
 & $\kappa$ & 2.89 & 2.71 & 0.20 & $-0.618$ \\
 & $\iota$ & 0.058 & 0.055 & 0.58 & $-0.349$ \\
 & $\delta$ & 0.627 & 0.647 & 0.61 & $+0.674$ \\
\bottomrule
\end{tabular}
\end{table}

\begin{table}[t]
\centering
\caption{\emph{In-sample (uncontrolled)} structural reliability signals on the
$103$-case golden set. Risk--coverage AURC is \emph{lower-is-better}; AUROC is
correct/incorrect triage. ``--'' marks quantities not of that type. These
case-level associations do \emph{not} survive capture-size control; see
\Cref{tab:controlled}.}
\label{tab:signals}
\small
\begin{tabular}{llccc}
\toprule
\# & Signal & Spearman $\rho$ ($p$) & AURC $\downarrow$ & AUROC \\
\midrule
-- & hand-tuned composite (baseline) & -- & 0.159 & 0.599 \\
1 & capture fragmentation $\kappa$-frac & $+0.250\ (0.011)$ & 0.132 & -- \\
1 & rank-mean(composite, separation) & -- & \textbf{0.129} & -- \\
2 & edge-Jaccard stability & $+0.269\ (0.019)$ & -- & -- \\
2 & LCC-Jaccard stability & $+0.315\ (0.006)$ & -- & -- \\
2 & composite $+$ conn.-robustness & -- & -- & \textbf{0.637} \\
3 & motif surprise ($z$, bigrams) & $-0.194\ (0.050)$ & -- & 0.358 \\
4 & community concentration & $+0.054\ (0.59)$ & -- & -- \\
5 & mean cited PageRank centrality & $+0.223\ (0.024)$ & -- & -- \\
\bottomrule
\end{tabular}
\end{table}

\begin{table}[t]
\centering
\caption{\emph{Size-controlled} re-analysis. Partial Spearman controls for
capture size, citation count, and mean citation gap; $p$ is a capture-level
permutation test; $q$ is Benjamini--Hochberg over all $14$ candidate
constructions ($n{=}103$ cases / $87$ captures, except the agreement metrics
which require a second opinion, $n{=}76$ / $63$). \emph{No signal survives
multiple-comparison correction}; $\Phimax$ excess-over-null is a
degree-preserving negative control.}
\label{tab:controlled}
\small
\begin{tabular}{lcccc}
\toprule
Signal & raw $\rho$ & partial $\rho\,|\,$size & cap-perm $p$ & BH $q$ \\
\midrule
fragmentation $\kappa$-frac        & $+0.250$ & $+0.138$ & $0.125$ & $0.29$ \\
LCC-Jaccard stability              & $+0.315$ & $+0.275$ & $0.026$ & $0.18$ \\
edge-Jaccard stability             & $+0.269$ & $+0.252$ & $0.025$ & $0.18$ \\
mean cited PageRank centrality     & $+0.223$ & $+0.032$ & $0.350$ & $0.54$ \\
connectivity-robustness            & $+0.171$ & $+0.070$ & $0.545$ & $0.64$ \\
motif surprise                     & $-0.194$ & $-0.048$ & $0.451$ & $0.57$ \\
$\Phimax$ excess over null (ctrl.) & $+0.041$ & $-0.070$ & $0.735$ & $0.79$ \\
\bottomrule
\end{tabular}
\end{table}

\begin{table}[t]
\centering
\caption{Microservice RCA on the Train Ticket chaos-injection corpus ($451$
cases). The reverse-PageRank prior localizes the root-cause service best, but it
is \emph{evidence-blind} (uniform restart): the alarm-seeded walk
(\texttt{PPR (from alarms)}) and the degree heuristic show the reverse prior's
edge is edge \emph{orientation}, not evidence-seeding. Forward PageRank is poor
because it walks \emph{away} from causes.}
\label{tab:openrca}
\small
\begin{tabular}{lcccc}
\toprule
Localizer & top-1 & top-3 & MRR & mean pctile \\
\midrule
random & 0.350 & 0.780 & 0.582 & 0.590 \\
degree & 0.685 & 0.953 & 0.816 & 0.787 \\
PageRank (forward) & 0.126 & 0.661 & 0.421 & 0.301 \\
\textbf{PageRank (reverse)} & \textbf{0.891} & \textbf{0.978} & \textbf{0.938} & \textbf{0.961} \\
closeness (from alarms) & 0.013 & 0.592 & 0.341 & 0.173 \\
PPR (from alarms) & 0.530 & 0.945 & 0.726 & 0.637 \\
\bottomrule
\end{tabular}
\end{table}

\begin{table}[t]
\centering
\caption{Public PetShop benchmark~\cite{hardt2024petshop} ($68$ issues),
node-level top-$k$ recall in PetShop's harness. Our relevance walk leads the
structure-only solvers on top-1 and is orientation-sensitive, but published
domain-native causal baselines are stronger on top-3---the honest limit of a
structure-only prior on metric-rich data.}
\label{tab:petshop}
\small
\begin{tabular}{lcc}
\toprule
Method & top-1 & top-3 \\
\midrule
degree (structural only)        & 0.28 & 0.54 \\
anomaly-only (no graph)         & 0.15 & 0.41 \\
relevance walk, reverse         & 0.06 & 0.27 \\
\textbf{relevance walk (ours)}  & \textbf{0.31} & 0.44 \\
\midrule
\multicolumn{3}{l}{\emph{published baselines~\cite{hardt2024petshop} (top-3 recall, range over scenarios):}}\\
graph traversal                 & --   & $0.57$--$1.00$ \\
CIRCA (causal)                  & --   & up to $1.00$ \\
counterfactual attribution      & --   & up to $0.86$ \\
\bottomrule
\end{tabular}
\end{table}

\section{Algorithms, Parameters, and Datasets}
\label{app:algo}

This appendix records the implementation details needed to reproduce every
reported number. All randomized procedures are seeded; the disease dataset is
obtained from its public source.

\paragraph{Evidence-graph construction (PCAP).}
Given a diagnosis citing frame set $C$, two frames $i,j\in C$ are joined iff
$|i-j|\le 1$ in capture order (a unit-width temporal bridge) or they belong to
the same endpoint conversation. The four order parameters of
\Cref{ssec:orderparams} are computed on the induced subgraph $G[C]$ by
union--find for components and direct edge counting for density.

\paragraph{Personalized PageRank.}
We solve \Cref{eq:ppr} with damping $d=0.85$ by power iteration to tolerance
$10^{-9}$ ($12$ iterations for the large sparse QMR graphs, up to $100$
elsewhere). The restart distribution is the seed set (the observed evidence) for
the evidence-seeded walks; the \emph{reverse-PageRank prior} reported as the
headline on the Train Ticket corpus instead uses a \emph{uniform} restart on the
reverse-transposed adjacency, so it ranks upstream sources without reading the
alarms (we report the alarm-seeded variant separately, \Cref{tab:openrca}). The
sparse implementation propagates mass only over realized edges, matching
\Cref{eq:ppr-series} truncated at the iteration count.

\paragraph{Noisy-OR oracle.}
We evaluate \Cref{eq:noisyor} with link reliabilities clamped to $[0.02,0.98]$
and per-finding leak $\ell_f=\max(\text{floor},\,0.3\,\hat p(f))$ estimated from
data, where $\hat p(f)$ is the empirical marginal of finding $f$. The synthetic
oracle uses the true generative parameters; the real-data oracle uses empirical
$\hat\Prob[f\mid d]$. Posteriors are computed in log-space for numerical
stability.

\paragraph{Synthetic percolation benchmark.}
Background mean degree $c=4$ (theory $\phi_c=1/c=0.25$), $L=4$ causal paths per
incident, decoy hubs carrying $+25$ excess degree clustered in a reserved region
disjoint from the causal paths, and $40$--$80$ independent trials per
$(\phi,N)$ point. Recoverability is the existence of an observed path from $R$ to
$A$; the empirical $\hat\phi_c$ is the location of the susceptibility peak.

\paragraph{Cross-domain datasets.}
Train Ticket chaos-injection corpus~\cite{zhou2021trainticket}: $451$
Chaos~Mesh fault-injection cases; spans and containers are canonicalized to
their parent service, yielding a mean of $5.1$ candidate services per case. PowerGraph~\cite{varbella2024powergraph}: IEEE-39/118/24 and
the UK grid, loaded from MATLAB v7.3 (HDF5) cascade archives; cascade labels
record lines tripped \emph{beyond} the initial outage. Disease--symptom
data~\cite{itachi2019dataset}: $41$ diseases, $131$ symptoms, $4{,}920$
patient records; edge weights are empirical $\Prob[\text{symptom}\mid
\text{disease}]$.

\paragraph{Statistical procedures.}
Paired pre/post comparisons use the Wilcoxon signed-rank test; monotone
association uses the Spearman rank correlation $\rho$ after removing undefined
entries; selective prediction uses the risk--coverage AURC and fixed-coverage
accuracy; ranking and triage use the Mann--Whitney AUC / AUROC. All metrics are
defined in \Cref{ssec:metrics}.

\end{document}